\documentclass[preprint, amsmath,amssymb,aps]{revtex4-1}%\documentclass[11pt]{article}

\usepackage{graphics}
\usepackage[dvips]{color}

\usepackage{graphicx}% Include figure files
\usepackage{dcolumn}% Align table columns on decimal point
\usepackage{bm,lineno}% bold math

\usepackage[normal]{subfigure}

\begin{document}
\linenumbers
\title{Entropy Production and Coarse Graining of the Climate Fields in a General Circulation Model}
\author{Valerio Lucarini [1,2]}
\email{Email: \texttt{valerio.lucarini@uni-hamburg.de}}
\author{Salvatore Pascale [1]}
\affiliation{\emph{[1]} Klimacampus, Meteorologisches Institut, Universit\"at Hamburg, Hamburg, Germany\\
\emph{[2]} Department of Mathematics and Statistics, University of Reading, Reading, UK}

\date{\today}

\begin{abstract}
We extend the  analysis of  the thermodynamics of the climate system by investigating the role played by processes taking place at various spatial and temporal scales through a procedure of coarse graining. We show that the coarser is the graining of the climatic fields, the lower is the resulting estimate of the material entropy production. In other terms, all the spatial and temporal scales of variability of the thermodynamic fields provide a positive contribution to the material entropy production. This may be interpreted also as that, at all scales, the temperature fields and the heating fields resulting from the convergence of turbulent fluxes have a negative correlation, while the opposite holds between the temperature fields and the radiative heating fields. Moreover, we obtain that the latter correlations are stronger, which confirms that radiation acts as primary driver for the climatic processes, while the material fluxes dampen the resulting fluctuations through dissipative processes. We also show, using specific coarse-graining procedures, how  one can separate the various contributions to the material entropy production coming from the dissipation of kinetic energy, the vertical sensible and latent heat fluxes, and the large scale horizontal fluxes, without resorting to the full three-dimensional time dependent fields. We find that most of the entropy production is associated to irreversible exchanges occurring along the vertical direction, and that neglecting the horizontal and time variability of the fields has a relatively small impact on the estimate of the material entropy production. The approach presented here seems promising  for testing climate models, for assessing the impact of changing their parametrizations and their resolution, as well as for investigating the atmosphere of exoplanets, because it allows for evaluating the error in the estimate of their thermodynamical properties due to the lack of high-resolution data. The findings on the impact of coarse graining on the thermodynamic fields on the estimate of the material entropy production  deserve to be explored in a  more general context, because they provide a way for understanding the relationship between forced fluctuations and dissipative processes in continuum systems.\end{abstract}

\maketitle

\section{Introduction}
%A bit of poema on non equilibrium thermodynamic and climate....
Along the lines of the  theoretical construction due to Lorenz \citep{Lor55,Lor67} of energy cycle of the atmosphere, the climate can be seen as a non-equilibrium multi-scale system, which generates entropy through a variety of irreversible processes \citep{Peix2,Goody00,Paul,Paul2,Luc10}, and transforms moist static energy into mechanical energy, as it features a positive spatio-temporal correlation between heating and temperature patterns, so that it can be represented schematically as a heat engine with a given efficiency \citep{Johnson,Lucarini}. For a given value of the external and internal parameters, the climate system achieves a steady state by balancing the input and output of energy and entropy with the surrounding environment \citep{Peix}. The large scale motions of the geophysical fluids are at the same time the result of the mechanical work produced by the climatic engine, and contribute to reducing the temperature gradients which make the energy conversion possible \citep{Stone,Peix2}. Obtaining a closure to this problem would be equivalent to developing a self-consistent theory of climate dynamics. 

Developing a comprehensive theory of  climate dynamics is one of the grand contemporary scientific challenges, also for its obvious environmental, social and economical relevance, and it is far from being an accomplished task \cite{Held05,schneider2006}. In recent years, the extraordinary developments of planetary sciences coming from the discovery of extra-solar planets and the ensuing need for understanding the properties of  atmospheric circulations realized under physical and chemical conditions very different from those of the Earth and of the other solar planets has provided further stimulation in this direction \cite{Seager}.   

%Recently, a line has been drawn connecting the climate engine's efficiency, the intensity of the Lorenz energy cycle, the entropy production of the climate system. In particular, the efficiency of the equivalent thermal machine sets also the proportionality between the internal entropy fluctuation of the system and the lower bound to entropy production by the fluid compatible with the 2nd law of thermodynamics. Such a bound is basically given by the entropy produced by the dissipation of the mechanical energy, whereas the excess of entropy production is due to the transport of heat down the gradient of the temperature field. Therefore, it has been possible to introduce a index of irreversibility of the system able to characterize the relative importance of these two processes \citep{Lucarini}.  

%The tools of phenomenological non-equilibrium thermodynamics \citep{Pri,Maz} seem well suited for providing powerful tools for addressing the long-standing problem of characterizing the structural properties of the climate system \citep{Saltzman}. The goal is to define a unifying framework for understanding the climate's variability, its feedbacks, and its large-scale processes, including the atmosphere-ocean coupling, the hydrological cycle, as well as understanding the mechanisms involved in climate phase transitions observed at the so-called tipping points, i.e. conditions under which catastrophes may occur for small variations in the boundary conditions or in the internal parameters of the system \citep{Lenton,Pal}. 

While the thermodynamic interpretation of the baroclinic disturbances, which provide the dominant contributions to the low-to-high latitudes heat transport, lies at the core of dynamical meteorology \citep{Holton}, and thermodynamics provides indeed the best framework for studying strong meteorological features like hurricanes \citep{Emm}, recent results suggest that the structural properties of the climate system \cite{Saltzman} and in particular its tipping points \citep{Lenton,Pal} can be effectively analyzed using the thermodynamic indicators developed in \citep{Lucarini}, with the efficiency and the entropy production providing the most interesting indicators \citep{LucFr,Boschi2013,LucAstr}. Moreover, recent studies have underlined that it is instead possible to define generalized climate sensitivities able to describe quite accurately the responses of thermodynamic quantities to changes in CO$_2$ concentration  \citep{GenSens}. % and to link such responses to variations in the intensity of processes, such as the hydrological cycle, which are much more commonly considered in the context of climate change literature.

%As these results address the global structural properties of the CS, they pave the way for a new, extensive exploration aimed at understanding the climate response under various scenarios of forcings, of atmospheric composition, and of boundary conditions. focused on the impacts on the thermodynamics of the CS of changes in the solar constant, with the ensuing analysis of the onset and decay of snowball Earth conditions (Lucarini et al., 2010a), and on those due to changing levels of CO2 concentration (Lucarini et al., 2010b).

%The climate  is  a far-from-equilibrium steady state system and  a variety of irreversible processes  take place  within it. The physical quantity which describes  the irreversibility (non-equilibrium)  of the climate system  is its entropy production rate  (\cite{ Maz, Pri}).   The interest in studying climate irreversibility has been renewed in the last decades by the Maximum Entropy Production conjecture (MEP) (originally proposed by \cite{Pal}, for a review see \cite{Mart}, \cite{MEP} or \cite{Klei09}).  Although the rigorous justification of such a conjecture is still missing (\cite{ Grgr,De2}) and the practical utility in applications  to climate models fairly limited (\cite{Goody07, KuFrKi08, Pascale2}), MEP has stimulated a reexamination of entropy production in the climate system (\cite{Peix, Goody00, Frae, Pascale}) and the development of a new thermodynamic perspective  which may be of great relevant for providing new reliable metrics and methods to study climatic problems (\cite{Salzmann, LucFr}).

Despite its relevance at theoretical level \citep{Pri,Maz}, traditionally, entropy production is not one of the first physical quantities climate modelers investigate when assessing the performance of a global climate model (GCM) or the response of the climate system to forcings. %Nonetheless, recently it has been shown that, starting from the theoretical formulations proposed in \citep{Goody00,Lucarini,Pascale2}, it is possible to provide a  simplified expression for the entropy production based only upon the radiative balances at surface and at the top of the atmosphere. Such expression allows for separating the contribution to entropy production coming from vertical and horizontal irreversible exchange processes and can be used for testing GCMs, and for estimating or constraining the entropy production (and other thermodynamical quantities) for planetary objects for which only coarse resolution data are available and model-assisted reconstruction is mostly temptative \citep{Luc10}. % One must keep in mind that from estimates of the entropy production it is possible to construct bounds on the energetics of the circulation, which is a property of paramount importance for the definition of the characteristics of planetary atmospheres \citep{Luc10}.  
One should note that the attention towards entropy production in the climate system and in climate modeling has been revived when several authors started proposing it as target function to maximize when tuning free or empirical parameters of approximate numerical models \citep{Lorenz,Klei06,KuFrKi08} or for getting good first order approximations of the climate state without resorting to long integrations \citep{Pal,Herb1}  This is the \textit{weak} or \textit{pragmatic} version of the so-called maximum entropy production principle (MEPP) \citep{MEP}, which, in its \textit{strong} form, proposes that any non-equilibrium systems adjust itself in order to maximize the entropy production; see \citep{Oz,MEP}. The MEPP theoretical foundations \citep{De,De2} have been  criticized both at theoretical level \citep{Grgr} and in terms of its geophysical applications \citep{Goody07,Pascale2}, so that weaker formulations are now mostly preferred \citep{DewEntr}. %, while at the same time there are attempts at using the entropy production of the system as an approximate Lyapunov functional for the dynamical system under investigation \citep{Herb2}.  
Recently, some authors have turned their attention on testing whether it is possible to propose a a variational principle for applies to another index of the irreversibility of the system, namely the rate of dissipation of kinetic energy \cite{Kleidon2013,Pascale13}. 

In this paper, we wish to investigate the entropy production of a climate model for studying, instead of large scale balances, its fluctuations at different temporal and spatial scales. Climate is a multi-scale system where dynamics takes place on vast range of interacting scales. The definition of parametrizations for unresolved scales is a major challenge of climate modeling and the proposal of closure theories connecting small and large scale properties is a major part of any attempt at formulating approximate theories for climate dynamics. The issue of understanding the impact of small scales on large scales and vice-versa, and of performing properly the upscaling and downscaling of a model's output is of great relevance also for intercomparing the performances of various versions of a given numerical model, or of a set of numerical model simulating the same system, differing for the adopted spatial and temporal resolution, and for comparing model data to observations. 

Our goal is manifold. One one side, we want to introduce a way to evaluate how the different scales of motion contribute to the overall entropy production of the climate system. This investigation, therefore, complements the investigation of how much energy is contained in the various scales of motion and of the energy fluxes across these scales. In order to achieve this goal, we consider the entropy budget of the FAMOUS GCM \cite{FAM} in standard, present climate configuration, taking advantage of the fact that it is one of the very few climate models where the entropy production diagnostics has been implemented and throughly tested \cite{Pascale, Pascale2}.  Starting from the output fields at the highest possible resolution given by the model (temporal resolution of one time step, and same spatial resolution of the actual numerical model), we perform a coarse graining in space and in time to the dynamical and thermodynamical fields appearing in the terms describing the entropy production of the system, and we test how the estimate of the entropy production changes when different coarse graining are applied to the data. We anticipate that we obtain that the coarser is the graining procedure, the lower is the estimate of the entropy production one obtains, as somehow intuitive. One must note that this is, in fact an obvious result when considering simple diffusive system, but not so obvious when fully nonlinear, multiphase systems are considered. We also obtain similar results when considering different degrees of longitudinal averaging of the fields, up to considering zonally averaged fields only. Our findings provide a way to assess how having low-resolution information about the dynamics of turbulent systems affects our ability to reconstruct its thermodynamical properties. Moreover, the procedure discussed in this paper allows to put on firmer ground the results proposed in \citep{Luc10} on the possibility of separating vertical and horizontal exchange processes as far as entropy production is concerned. Finally, we can study in detail the relationship between two apparently equivalent ways of computing the entropy production proposed in \citep{Goody00}. 

This paper is structured as follows. In section \ref{budget} we  briefly recapitulate some definitions and equations relevant for setting the problem of computing the entropy production of the climate system we explain how to perform such a calculation in a climate model. In section \ref{scale} we explain what we mean precisely by coarse graining of the data and describe how it is actually implemented in the model's output. We also provide some conjectures what will be discussed in later in the paper. In section \ref{results} we present our results. We first describe the impact of performing coarse graining on time alone, thus exploring the range between time step data and long term averaged data, and then we extend our analysis to the space domain, showing how performing zonal, horizontal, and mass-weighted averaging over the output data impacts the obtained estimate of the entropy production. In section \ref{conclu} we present our conclusions and perspective for future works.  In appendix \ref{app} we present some theoretical arguments on a simple diffusive system for clarifying the meaning of the results obtained from the data analysis. 

\section{Climate entropy budget}\label{budget}

Following \citep{Pri,Maz}, for any system it is possible to decompose the rate of change of its entropy $\mathrm{d}S/\mathrm{d}t$ as $\mathrm{d}S/\mathrm{d}t = \mathrm{d}_eS/\mathrm{d}t+\mathrm{d}_iS/\mathrm{d}t$, where the first term is called the external and the second term is the internal contribution to the entropy budget. The external contribution corresponds to the entropy flux through the boundaries of the system whereas the internal entropy production is associated with the irreversible processes taking place in the system. The second law of thermodynamics imposes that the internal entropy production has to be nonnegative at all instants, so that $\mathrm{d}_iS/\mathrm{d}t\geq 0$. When a statistically steady state is achieved, the external and internal entropy production have to balance each other so that the total rate of entropy change is zero: we have $\overline{\mathrm{d}S/\mathrm{d}t} = 0 \rightarrow \overline{\mathrm{d}_iS/\mathrm{d}t}=-\overline{\mathrm{d}_eS/\mathrm{d}t}\geq 0,$ where the overline indicates averaging over a long time interval compared to the internal scales of the system. The  previous expression means that a non-equilibrium system generates on the average a positive amount of entropy through irreversible processes, and such excess of entropy is expelled at the boundaries. Non-equilibrium is maintained if the system is in contact with more than one reservoir with given temperature and/or chemical potential \citep{Gallavotti}. Of course, if the system is at equilibrium, the previous inequality becomes an equality, as in the long run the system reaches an homogeneous state of maximum entropy and no additional entropy is generated. 

In the climate system two rather different set of processes contribute to the total entropy production \citep{Peix2,Goody00}. The first set of processes are responsible for the irreversible thermalisation of the photons emitted near the Sun's corona at roughly 5700 K at the much lower temperatures, typical of the Earth's climate. This contributes for about 95\% of the total average rate of entropy production for our planet, which is about 0.90 $W$ $m^{-2}$ \citep{Peix2,Goody00}.  The remaining contribution is due to the processes responsible for mixing and diffusion inside the fluid component of the Earth system, and for the dissipation of kinetic energy due to viscous processes. This constitutes the so-called material entropy production, and is considered to be the entropy related quantity of main interest as far as the properties of the climate system are concerned. See \citep{Pascale2} for an extensive discussion of this issue a careful estimate of its value in two climate models, including the one used in this study.

When separating the entropy budget for radiation and for the fluid part of the climate system, and taking long term averages, one can derive the following equation \cite{Johnson,Goody00}:
\begin{equation}
\int_V d^3\mathbf{x} \hspace{3mm}\left[\overline{\left(\frac{\dot{q}_{rad}}{T}\right)}+\overline{ \dot{s}_{mat}}\right]    =0
\label{balance}
\end{equation}
where the  integral is over the whole volume $V$ of the climate system, $\dot{q}_{rad}$ is the radiative heating rate, $\dot{s}_{mat}$ is the instantaneous specific rate of material entropy production due to irreversible processes involving the climatic fluid, and $T$ the  temperature field. The following expression is usually adopted for $\dot{s}_{mat}$ \cite{Peix2,Johnson,Lucarini,Klei09}:
\begin{equation}
\dot{s}_{mat}=\frac{\epsilon^2}{T}+\mathbf{F}_{SH}\cdot\nabla{\left(\frac{1}{T}\right)}+\mathbf{F}_{LH}\cdot\nabla\left(\frac{1}{T}\right)
\label{dir}
\end{equation}
where $\epsilon^2$ the specific dissipation rate of kinetic energy,  $\mathbf{F}_{SH}$ the turbulent sensible heat flux , $\mathbf{F}_{LH}$ the turbulent latent heat flux, where by turbulent we mean not related to large scale advection due to winds, which is in principle reversible. Romps \cite{Romps} refers to the representation of the entropy production given by Eq. (\ref{dir}) as resulting from the bulk heating budget, because water is treated mainly as a passive substance, while processes such as irreversible mixing of water vapor are altogether ignored. More detailed description of the moist atmosphere have led to a 
consistent treatment of the entropy generated by the various processes accounting for hydrological cycle
these processes \cite{Goody00,Paul,Paul2,Romps}. Apparently, though, the overall effect of hydrological cycle-related entropy production is captured quite well using Eq. (\ref{dir}) \citep{Goody00,Luc10,Pascale}.  

Integrating the term $\dot{s}_{mat}$ in Eq. (\ref{balance}) over the volume $V$ of the climate system and taking a long-term average, we obtain the average rate of material entropy production:
\begin{equation}
\overline{\dot{S}_{mat}}=\int_V d^3\mathbf{x} \hspace{3mm} \overline{ \dot{s}_{mat}}=\overline{\dot{S}_{mat}^{dir}},
\label{dir_ind}
\end{equation}
which gives the so called \textit{direct formula} for the material entropy production. Using  Eq. (\ref{balance}), we derive an equivalent expression involving radiative heating rates only:
\begin{equation}
\overline{\dot{S}_{mat}}=-\int_V  d^3\mathbf{x} \hspace{3mm} \overline{\left(\frac{\dot{q}_{rad}}{T}\right)}=\overline{\dot{S}_{mat}^{ind}},
\label{ind_ind}
\end{equation}
which is the \textit{indirect formula} for computing the average rate of entropy production, where obviously $\overline{\dot{S}_{mat}^{dir}}=\overline{\dot{S}_{mat}^{ind}}=\overline{\dot{S}_{mat}}$.  Equation (\ref{balance}) provides an intimate link between the radiative fields and the material flow properties inside the climate system. Moreover, Eq. (\ref{ind_ind})  is very powerful because it permits to work out the average rate of material entropy production by considering only the optical properties of the fluid. Pascale et al. \cite{Pascale} showed using the climate model FAMOUS adopted in this study that $\overline{\dot{S}_{mat}^{dir}}$ and $\overline{\dot{S}_{mat}^{ind}}$ agree up to an excellent degree of precision (within $1\%$). Since Pascale et al. \citep{Pascale} used the approximate expression (\ref{dir}) for the specific material entropy production, they also confirmed that, indeed, at all practical purposes using such simplified representation of the irreversibility associated to the hydrological cycle is appropriate.

\subsection{Entropy diagnostics in Climate Models}
\label{GCM}
In an actual climate model the implementation of entropy diagnostics faces some difficulties, both at theoretical level and in terms of practical implementation of the entropy-related diagnostics.   A theoretical difficulty is that, as evidenced in \citep{LucariniRagone}, many state-of-the-art climate models features an inconsistent energetics, such that when all parameters are held fixed and the system reaches a steady state, the long-term average of the energy budget at the top of the atmosphere (TOA), which is the only boundary of the climate system, is unexpectedly biased with respect to the vanishing long-term average one should expect to observe. Interestingly, all biased models feature a positive energy budget at TOA, which implies that the time averaged outgoing long wave radiative flux is smaller than the net incoming shortwave flux. This fact implies that there must be a positive definite spurious sink of energy somewhere inside the system. More specific analyses make clear that such spurious sinks are related to the imperfect closure of the hydrological cycle \citep{Liepert} and to the inconsistent treatment of the dissipation of kinetic energy, which is not entirely (or at all) fed back into the system as thermal energy \citep{LucariniRagone}. Such  inconsistencies at smallspatial and temporal  scales impact large scale, long term  climatic properties. As a result, climate models are biased cold, taking into consideration that the Earth emits approximately as black body, or feature negative biases in the planetary albedo, or both. Moreover, since the biases are related to climate processes, they are climate-dependent, and so hard to control a posteriori via removal of anomalies. In terms of entropy production, an energy bias of the order of $1$ $W$ $m^{-2}$ causes a bias in the entropy production of about $4\times10^{-3}$ $W$ $m^{-2}$ $K^{-1}$, which is comparable with the range of estimates of material entropy production given by various climate models \citep{Pascale,Luc10}. The FAMOUS model we use in this study features minor inconsistencies in terms of closure of the energy budget (the bias is smaller than $0.1$ $W$ $m^{-2}$, so that the problem exposed here does not affect significantly our results (see discussion later).

Moreover, in a climate model it is hard to deal directly with Eq.~(\ref{dir}) because material turbulent fluxes are evaluated through parametrizations of unresolved processes. The corresponding routines in the numerical code do not give as outputs heat fluxes. On the  other hand the heating rates (i.e. the divergence of the heat fluxes)  are easily  diagnosed for these unresolved processes. %Nonetheless, we can easily convince ourselves that using Gauss' theorem we can overcome this difficulty. %
%
%A heating rate can be written in terms of the convergence of a flux (say e.g. $\mathbf{F}_{LH}$) as follows:
%\begin{equation}
%\int_V \frac{-\nabla\cdot\mathbf{F}_{LH}}{T} d^3\mathbf{x}=\int_V \mathbf{F}_{LH}\cdot\nabla\left(\frac{1}{T}\right)-\int_{\partial V} \frac{\mathbf{F}_{LH}}{T}\cdot\mathbf{n},\,dA
%\label{convergence}
%\end{equation} 
%over a volume $V$ with boundary $\partial V$. If we integrate over the whole climate system, $\partial V$ is the top-of-the-atmosphere  and the bottom of the solid Earth. 
Neglecting the geothermal flux from the inner Earth and  noting that at the top-of-the-atmosphere we have only radiative fields, %we can  we conclude that the boundary term is zero.  Hence, when the whole climate system is taken into account, 
using Gauss' theorem, the material entropy production can be worked out by considering all the material diabatic heating rates, as shown in \cite{Pascale}:
\begin{equation}
\overline{\dot{S}^{dir}_{mat}}=\int_V\mathrm{d}^3\mathbf{x}\overline{\left(\frac{\epsilon^2}{T}\right)}-\overline{\left(\frac{\nabla\cdot \mathbf{F}_{SH}}{T}\right)}-\overline{\left(\frac{\nabla\cdot \mathbf{F}_{LH}}{T}\right)}=\int_V d^3\mathbf{x}\overline{\left(\frac{\dot{q}_{mat}}{T}\right)}
\label{dirfinale}
\end{equation}
In this paper we refer to the entropy budget of the FAMOUS GCM  \cite{FAM}  which has been studied in detail  by \cite{Pascale, Pascale2}. Lets first focus on the evaluation of $\overline{\dot{S}_{mat}^{dir}}$. Different processes contribute to the entropy production terms described in Eq.(\ref{dir}): the heating rates are calculated as output of many different parametrization routines describing the unresolved processes in the various subdomains of the climate system (atmosphere, ocean, soil, cryosphere): 
\begin{itemize}
\item Entropy production due to dissipation of kinetic energy, $\overline{\dot{S}_{KE}}$,  defined as:
\begin{equation}
\overline{\dot{S}_{KE}}=\int  d^3\mathbf{x} \overline{\left(\frac{\epsilon^2}{T}\right)} .
\label{KE}
\end{equation}
In FAMOUS and, in general, in most  climate models, the kinetic energy is dissipated mainly through four parametrized processes: the turbulent stresses occurring at the boundary layer, which extract kinetic energy from the free atmosphere, the gravity wave drag, which dissipates kinetic energy in the upper atmosphere, atmospheric convective processes, and small scale turbulence, which is represented by the horizontal momentum hyperdiffusion (which serves also the purpose of increasing the numerical stability of the model). In FAMOUS only the atmosphere contributes to this part of the entropy production. This is a reasonable approximation because the dissipation of kinetic energy occurring in the atmosphere is about two orders of magnitude larger than that occurring in the ocean \cite{Peix2,VonStorch}.
\item Entropy production due to irreversible transfer of sensible and latent heat via turbulent fluxes, $\overline{\dot{S}_{heat}}=\overline{\dot{S}_{SH}}+\overline{\dot{S}_{LH}}$, defined as: 
\begin{equation}
\overline{\dot{S}_{heat}}=\int  d^3\mathbf{x}\left[ -\overline{\left(\frac{\nabla \cdot \mathbf{F_{SH}}}{T}\right)}-\overline{\left(\frac{\nabla \cdot \mathbf{F_{LH}}}{T}\right)}\right]=\overline{\dot{S}_{SH}}+\overline{\dot{S}_{LH}}
\label{HEAT}
\end{equation}
%where we have separated the contributions coming from irreversible sensible and latent heat transfer. 
The boundary layer scheme contributes to the entropy production due to irreversible sensible and latent heat transfer in the four subdomains of the climate system, as it couples them through exchanges of sensible heat and water vapour; other parametrized processes contributing to $\overline{\dot{S}_{SH}}$ and $\overline{\dot{S}_{LH}}$ are  atmospheric convection and the condensation and evaporation of water in the atmosphere, as determined by the clouds and precipitation parametrization schemes. Instead, processes contributing only to $\overline{\dot{S}_{SH}}$ are the oceanic convection, the small scale turbulent mixing of temperature described by hyperdiffusion, and the mixing occurring inside the ocean associated to small scale eddies and in the mixed layer. %One must note a basic difference between the contribution to entropy production due to sensible and latent heat irreversible transfer is that the former is eminently a process where heat is absorbed and released in the same atmospheric column. In the case of latent heat, winds can transport away water vapor so that condensation of a water vapor parcel can take place at large horizontal distances from the region where evaporation had taken place. % The closure of such mass transport through river and ocean transports is the essential element of the hydrological cycle.
\end{itemize}

Table \ref{proce} provides a synthetic outline of which routines describing unresolved processes contribute to the various terms of the material entropy production in each climatic subdomain. Therefore, in practice, we compute $\overline{\dot{S}_{mat}^{dir}}$ as follows:
\begin{equation}
\overline{\dot{S}_{mat}^{dir}}=\sum_k\sum_c\int_{V_c}  d^3\mathbf{x} \overline{\left(\frac{\dot{q}_k^c}{T}\right)} 
\label{matter}
\end{equation}
where $\dot{q}_k^c$ is the local instantaneous heating rate occurring in the subdomain $V_c$  due to the process $k$.% As described in Table \ref{proce}, we emphasize that $\dot{q}_k^c$ can contribute to $\epsilon^2$, $-\nabla \cdot \mathbf{F_{LH}}$, or $-\nabla \cdot \mathbf{F_{SH}}$.

% terms of the entropy sources (\ref{tend}) the global integral rate of material entropy production is written as \cite{Pascale}:
% - where the index refers to diffusion, numerical hyperdiffusion, condensation, atmospheric and oceanic convection, boundary layer transport of sensible heat and water vapour, evaporation of falling rain, dissipation of kinetic energy) and the integrals are extended over the volume of the climate subsystems $V_c$.  $S_{mat}^{ind}$ is worked out similarly  according to Eq.~\ref{ind_ind} 

The evaluation of $\overline{\dot{S}_{mat}^{ind}}$ is much easier because the heating rates are readily available from the radiation scheme, which affects all the subdomains $c$ of the climate system: 
\begin{equation}
\overline{\dot{S}_{mat}^{ind}}=-\sum_c\int_{V_c}  d^3\mathbf{x}\left[ \overline{\left( \frac{\dot{q}_{sw}^c}{T}\right)} +\overline{\left(\frac{\dot{q}_{lw}^c}{T} \right)}\right] 
\label{radiation}
\end{equation}
where we have divided the contribution $\dot{q}_{sw}$ coming from the shortwave radiation, which is only absorbed (and scattered),  inside the climate systems, so that $\dot{q}_{sw}\geq 0$, from the contribution $\dot{q}_{lw}$ coming from the longwave radiation, which instead is scattered, absorbed, and emitted, and is the sole responsible for the radiative cooling.

 \section{Coarse-Graining of the Entropy Production Terms: Definitions and Some Conjectures}\label{scale}

The entropy budget is estimated using space and time integrals of the ratio between the local heating term and the local temperature. In many cases, either because  we need to compress data or because climatological database only contain certain time or spatially averaged data, we have to deal with coarse grained data  for the heating rate $\langle q(\mathbf{x}, t)\rangle_v^{\tau}$, and for the temperature $\langle T(\mathbf{x}, t)\rangle_v^{\tau}$, where $\tau$ refers to the time scale of the temporal averaging operation, and $v$ refers to the set of stencil regions $v(x)$ centered over $x$ over which (mass-weighted) spatial averaging is performed:  
\begin{equation}
\langle X(\mathbf{x}, t)\rangle_v^{\tau}=\frac{1}{\tau \mu(v,t)}\int_{v}  d^3\mathbf{y} \int_{-\tau/2}^{\tau/2}d \sigma X(\mathbf{x+y},t+\sigma)
\label{cgraining}
\end{equation}
%  from time means of terms having the form 
%\begin{equation}
%\int  d^3\mathbf{x} \frac{q_k(\mathbf{x},t)}{T_k(\mathbf{x},t)}
% \label{ratio}
% \end{equation}
% where $q_k(\mathbf{x}, t)$ is the  heating at the time $t$ and at the point $\mathbf{x}$ associated with a diabatic process $k$, $T_k(\mathbf{x}, t)$ the  temperature at which the diabatic  heat $q_k$  is  added (see for example \cite{Frae,Pascale}) and the integral is usually over the climate system domain. 
where $\mu(v(x),t)=\int_{v(x)}  d^3\mathbf{x} $ is the mass contained in the stencil $v(x)$ at time $t$. Mass-weighting is the natural choice in climate models as hydrostatic approximation is almost invariably used and vertical coordinates are expressed to a very good approximation in terms of pressure levels.
Since the integrands in Eqs. (\ref{ind_ind}) and (\ref{dirfinale}) are nonlinear, we obviously have that for every $\tau$ and $v$:
\begin{align}
\overline{\dot{S}^{dir}_{mat}}=\int_V d^3\mathbf{x}\overline{\left(\frac{\dot{q}_{mat}}{T}\right)} &\neq \int_V d^3\mathbf{x}\overline{\left(\frac{\langle \dot{q}_{mat}\rangle_v^\tau}{\langle T\rangle_v^\tau}\right)}=\overline{\langle \dot{S}^{dir}_{mat} \rangle_v^\tau},\label{c1}\\
\overline{\dot{S}^{ind}_{mat}}=-\int_V d^3\mathbf{x}\overline{\left(\frac{\dot{q}_{rad}}{T}\right)} &\neq - \int_V d^3\mathbf{x}\overline{\left(\frac{\langle \dot{q}_{rad}\rangle_v^\tau}{\langle T\rangle_v^\tau}\right)}=\overline{\langle \dot{S}^{ind}_{mat} \rangle_v^\tau}.\label{c2}
\end{align}
Moreover, while as discussed before $\overline{\dot{S}^{dir}_{mat}}=\overline{\dot{S}^{ind}_{mat}}$, there is no \textit{a priori} reason to expect that $\overline{\langle \dot{S}^{dir}_{mat} \rangle_v^\tau}$ and $\overline{\langle \dot{S}^{ind}_{mat} \rangle_v^\tau}$ have the same value. Finally, we have that up to first order: 
\begin{align}
\overline{\dot{S}^{dir}_{mat}}-\overline{\langle \dot{S}^{dir}_{mat} \rangle_v^\tau}= \Delta\left[\overline{S^{dir}_{mat}} \right]_v^\tau \simeq - \int_V d^3\mathbf{x}\overline{\frac{\Delta\left[\dot{q}_{mat}\right]_v^\tau \Delta\left[T\right]_v^\tau}{\left[\langle T\rangle_v^\tau\right]^2}}\label{d1}\\
\overline{\dot{S}^{ind}_{mat}}-\overline{\langle \dot{S}^{ind}_{mat} \rangle_v^\tau}= \Delta\left[\overline{S^{ind}_{mat}} \right]_v^\tau \simeq  \int_V d^3\mathbf{x}\overline{\frac{\Delta\left[\dot{q}_{rad}\right]_v^\tau \Delta\left[T\right]_v^\tau}{\left[\langle T\rangle_v^\tau\right]^2}}\label{d2}%\overline{S^{ind}_{mat}}-\overline{\langle S^{ind}_{mat} \rangle_v^\tau}=\Delta\left[\overline{S^{ind}_{mat}} \right]_v^\tau \simeq \int_V d^3\mathbf{x}\overline{\left(\frac{\Delta\left[\dot{q}_{rad}\right]_v^\tau \Delta\left[T\right]_v^\tau}{\left[\langle T\rangle_v^\tau}\right]^2\right)}\label{d2}
\end{align}
where $\Delta\left[X\right]_v^\tau=X-\langle X\rangle_v^\tau$. It is natural to interpret  $\overline{\langle \dot{S}^{ind}_{mat} \rangle_v^\tau}$, $\overline{\langle \dot{S}^{dir}_{mat} \rangle_v^\tau}$ as the contribution to the entropy production due to irreversible processes occurring on scales large than what described by $\tau$ and $v$. Consequently, $\Delta\left[\overline{\dot{S}^{ind}_{mat}} \right]^{\tau}_v$, $\Delta\left[\overline{\dot{S}^{dir}_{mat}} \right]^{\tau}_v$ in Eqs. (\ref{d1})-(\ref{d2}) can be interpreted as the contributions to the entropy production given by the material flows (Eq. (\ref{d1})) and radiative fluxes (Eq. (\ref{d2})) with variability confined below the spatial scale given by $v$ and by the time scale given by $\tau$.

Equations (\ref{c1})-(\ref{d2}) address practical questions such as: what is the error related to remapping the output of a climate model to a new resolution in space and time? How do diurnal, seasonal and interannual variability  and how different spatial structures (midlatitude cyclones, equator-pole contrasts, longitudinal asymmetries due to ocean-land contrasts, etc) affect the entropy budget?  How should we proceed to compare the estimates of material entropy production from models with different resolutions? Moreover, we need to understand whether it is more accurate to obtain estimates of the material entropy production from coarse grained fields of the radiative heating rates or of the material heating rates, which can be used for the indirect  or direct formula for the entropy production, respectively. These issues may be sequentially investigated by filtering $q(\mathbf{x},t)$ and $T(\mathbf{x},t)$ over the associated time- and space- scales. 

%Note that this matter becomes extremely relevant when we consider the possibility of estimating the material entropy production for planetary systems where our ability of observing or modeling at high spatial and temporal resolution is limited. 

When we perform the coarse graining given in Eq. (\ref{cgraining}) to the thermodynamic variables and estimate the entropy production, we discount for the mixing processes occurring below the chosen spatial and time scales. Therefore, one expects that $\overline{\langle \dot{S}^{ind}_{mat} \rangle_v^\tau}$, $\overline{\langle \dot{S}^{dir}_{mat} \rangle_v^\tau}\geq 0$ and $\Delta\left[\overline{\dot{S}^{ind}_{mat}} \right]^{\tau}_v,\Delta\left[\overline{\dot{S}^{dir}_{mat}} \right]^{\tau}_v\geq 0$ for all choices of $\tau$ and $v$.  Moreover, it seems natural to conjecture that if, given a model output, we choose a coarser graining, we should obtain a lower estimate of the entropy production, because we neglect the impact of a larger set of irreversible processes. In other terms, we should have that $\Delta\left[\overline{\dot{S}^{dir}_{mat}} \right]_{v_{1}}^{\tau_{1}}\geq\Delta\left[\overline{\dot{S}^{dir}_{mat}} \right]_{v_{2}}^{\tau_{2}}$ (or $\overline{\langle \dot{S}^{dir}_{mat}\rangle_{v_{1}}^{\tau_{1}}}\leq \overline{\langle \dot{S}^{dir}_{mat}\rangle_{v_{2}}^{\tau_{2}}}$) and $\Delta\left[\overline{\dot{S}^{ind}_{mat}} \right]_{v_{1}}^{\tau_{1}}\geq\Delta\left[\overline{\dot{S}^{ind}_{mat}} \right]_{v_{2}}^{\tau_{2}}$ (or $\overline{\langle \dot{S}^{ind}_{mat}\rangle_{v_{1}}^{\tau_{1}}}\leq \overline{\langle S^{ind}_{mat}\rangle_{v_{2}}^{\tau_{2}}}$) if $\tau_2\leq \tau_1$ and $v_2\subset v_1$.  Let's see how to interpret the inequalities $\Delta\left[\overline{\dot{S}^{ind}_{mat}} \right]^{\tau}_v$, $\Delta\left[\overline{\dot{S}^{dir}_{mat}} \right]^{\tau}_v\geq 0$ using the r.h.s. of Eqs. (\ref{d1})-(\ref{d2}):\begin{itemize}
\item The inequality $\Delta\left[\overline{\dot{S}^{ind}_{mat}} \right]^{\tau}_v\geq 0$ can be interpreted as the fact that at all time and space scales, there is on the global average a \textit{positive} correlation between the anomalies of radiative heating and the anomalies of temperature. This expresses the basic fact that the climate system is driven by radiative forcings, in the first place. Hence, this term refers to the \textit{response of the system to the external forcing}. Note that the inequality holds despite the the strong negative correlation between temperature anomalies and long wave heating rate anomalies due to the Boltzmann feedback. 
\item The other inequality $\Delta\left[\overline{\dot{S}^{dir}_{mat}} \right]^{\tau}_v\geq 0$, instead, implies that at all time and space scales on the average there is a \textit{negative} correlation between the anomalies of heating due to convergence of material heat fluxes and anomalies of temperatures. This relation, instead, expresses the fact that   temperature anomalies are damped by the geophysical flows, and this terms refers to the \textit{dissipation} occurring inside the system at all scales. 
\end{itemize}
In other terms, these conjectured inequalities correspond to the well-known fact that the climate is 1. forced by anomalies in the radiative forcing, and 2. the atmospheric and oceanic circulations reduce the resulting temperature gradients. Climate processes are related in such a way at all scales, \textit{in the average}, i.e. when space and time averages are considered. Obviously, locally in space and/or in time one can get, e.g. positive temperature fluctuations and, at the same time, a positive heating due to latent heat release and sensible heat convergence (e.g. tropical troposphere). Such processes can be positively correlated in time for some locations, but this must come at the expenses of negative correlations dominating elsewhere in the globe.

As the primary driving of climate is indeed the radiative forcing, while the fluid flows tend to dampen the resulting temperature gradients through instabilities, Therefore, one expects that the correlations between temperature and heating fields are stronger when considering the radiative fields as sources of heating.  In other terms, the convergence of heat due to geophysical flows are neither strong nor fast enough to counter exactly the radiative forcing at all scales. As an example, one may consider the fact that the radiative-convective equilibrium is typically baroclinically unstable in the mid-latitudes, and, indeed, baroclinic disturbances reduce the North-South temperature gradient by transporting heat from South to North, but cannot reduce it to zero.  Taking into consideration Eqs. (\ref{d1})-(\ref{d2}), the different role - forcings vs. dampening - of the convergence of the radiative fluxes vs. material turbulent fluxes leads us to proposing an additional inequality. We conjecture that $$\Delta\left[\overline{S^{ind}_{mat}} \right]^{\tau}_v\geq \Delta\left[\overline{S^{dir}_{mat}} \right]^{\tau}_v \qquad \forall \tau,v, $$ from which, since $\overline{S^{dir}_{mat}}=\overline{S^{ind}_{mat}}$, we derive the following inequality $$\overline{\langle S^{dir}_{mat} \rangle_v^\tau}\geq \overline{\langle S^{ind}_{mat} \rangle_v^\tau}, \qquad \forall \tau,v.$$

\section{Results}\label{results}
%\subsection{Temporal Coarse Graining}
We first discuss briefly how the coarse graining operation is performed in practice. Let us consider a steady-state climate simulation lasting for  a time period $L$ (in our case $L=50$ years), which we divide it in $N$ sub-intervals $\tau=L/N$, where $\tau = M \times \mathrm{d}t$, where $\mathrm{d}t$ is the model's time step ($1$ $h$ in our case). %In the following we assume that the fields are continuous functions of the time, however we must bear in mind that they are may be known at most at each model time step ($1$ hour for atmospheric fields and $12$ hours for oceanic fields).
The horizontal resolution is specified by regular grids with angle resolution of $5^{\circ}\times 7.5^{\circ}$ lat-lon), while in the vertical we have 11 levels for the atmosphere, 20 oceanic levels, and 3 land surface levels \citep{FAM}. Therefore, we subdivide the domain $V$ of integration into $Q$ subdomains $v_q$, $q=1,\ldots,Q$, each containing (in the bulk of the model's domain) $R$ grid points. Given an intensive thermodynamic field $X(\mathrm{x}_k,t_j)$, for  $n=1,\ldots,N$ and $q=1,\ldots,Q$, we define its coarse grained version as: 
\begin{equation}
\langle X(q,n)\rangle_v^{\tau}=\frac{1}{\tau \mu(v_q,n)} \sum_{j=R(q-1)+1}^{Rq} \sum_{k=M(n-1)+1}^{Mn}\mathrm{d}t\mu(\mathbf{x}_j,\sigma_k)X(\mathbf{x}_j,\sigma_k),
\label{cgraining2}
\end{equation}
where $\mu(\mathbf{x}_j,\sigma_k)=\nu(\mathbf{x}_j) \rho(\mathbf{x}_j,\sigma_k)$ is the mass contained in the grid box centered around $\mathbf{x}_j$ of volume $\nu(\mathbf{x}_j) $ at time $\sigma_k$ and $\mu(v_q,n)$, correspondingly, is the time averaged (for time ranging from $t_{M(n-1)+1}$ and $t_{Mn}$ ) mass contained in the domain $v_q$ of volume $\nu(v_q)$.  Therefore, our estimate of the coarse grained value of the material entropy production is:
\begin{equation}
\overline{\langle \dot{S}^{dir}_{mat} \rangle_v^\tau}=  \frac{1}{N}\sum_{i=1}^{Q}  \sum_{l=1}^{N} \nu(v_i) \frac{\langle  \dot{q}_{mat}(i,l)\rangle_v^{\tau}}{\langle T(i,l)\rangle_v^\tau}\label{cgrainingdir}
\end{equation}
for the so-called direct formula, and:
\begin{equation}
\overline{\langle \dot{S}^{ind}_{mat} \rangle_v^\tau}= - \frac{1}{N}\sum_{i=1}^{Q}  \sum_{l=1}^{N} \nu(v_i) \frac{\langle  \dot{q}_{rad}(i,l)\rangle_v^{\tau}}{\langle T(i,l)\rangle_v^\tau} \label{cgrainingind}
\end{equation}
for the so-called indirect formula. These formulas are the discretized versions of Eq. (\ref{c1}) and Eq. (\ref{c2}), respectively. Obviously, the discrete versions of the exact formulas for the material entropy production $\overline{\dot{S}_{mat}^{dir}}$ and $\overline{\dot{S}_{mat}^{ind}}$ are obtained by setting in Eq. (\ref{cgraining2}) $R=M=1$, {i.e.}, taking the model outputs at the highest possible resolution. The processes occurring in the interior of the ocean and below the first soil level, as these contributions have been shown to be entirely negligible in terms of entropy production \citep{Pascale2}, and so are discarded.

If we choose a given spatial resolution of our data and we consider different values of $M$, we test how applying temporal coarse graining impacts the estimates of the material entropy production. Instead, if we change the shape of the stencil $v$ and/or the number of points $R$ while keeping $M$ fixed, we investigate the impact of changing the  spatial coarse graining scheme.  Obviously, we cannot capture the contributions to the material entropy production due to irreversible processes taking place over timescale shorter than the model timestep and over space scales smaller than the model resolution. It is not clear, given a specific model's settings, how relevant these could be, and, indeed, the only way to find this out is to alter the model's resolution. This procedure may have relevance in terms of model tuning, as one could decide to change a model's parameter when altering its resolution in such a way to keep the entropy production constant. 

%Indeed, these may significantly increase the overall budget but they are not represented in the model. So what we discuss here is the entropy production budget due to the large-scale of climate.

%The coarse grained value of  the and temperature over a time length $\tau$ at the time $t_j=(j-1/2)\tau$  (with $j=1,\ldots N$) is
%\begin{equation}
%\overline{ q(\mathbf x, t_j)} ^{\tau}=\frac{1}{\tau}\int_{t_j-\frac{\tau}{2}}^{t_j+\frac{\tau}{2}} q(\mathbf{x},t) dt, \quad \overline{T(\mathbf{x}, t_j)}^{\tau}=\frac{1}{\tau}\int_{t_j-\frac{\tau}{2}}^{t_j+\frac{\tau}{2}} T(\mathbf{x}, t_j) dt.
%\label{meaning}
%\end{equation}
\subsection{Temporal coarse graining}
We start our investigation by performing coarse graining exclusively on time. We then analyze a long, steady state model's run lasting $50$ years with a model's timestep of 1 hour, and consider 1 year as long-term averaging time. We use the following values for $\tau$: $1$ hour  (model timestep, $N=1$), $6$ hours ($N=6$) , $12$ hours ($N=12$), $1$ day ($N=24$), $2$ days ($N=48$), $5$ days ($N=120$), $10$ days ($N=240$), $15$ ($N=360$) days, $1$ month ($N=720$), $3$ months ($N=2160$), $6$ months ($N=4320$), $1$ year ($N=8640$). We then collect the 50 1-year averaged value of the coarse grained material entropy production and compute the mean and standard deviation for the 50 data we have. 
%These data are collected to analyze the properties of the climate system for scales ranging up to 1 year. 
Moreover, we consider longer averaging periods - $5$ years, $10$ years, and $50$ years, and take in these cases $\tau$ equal to the averaging time, so that $N=43200$, $N=86400$, and $N=432000$ in the $\tau=5$, $10$, and $50$ years case, respectively,  thus spanning in total more than 5 orders of magnitude for $N$. We then compute for the coarse-grained estimates of the material entropy production the ten $5-$year averages and the five $10-$year averages, and compute the mean and standard deviation, plus the unique value referred to the $50-$ year average. The statistics for such large values of $\tau$ are extremely stable.  

\begin{figure}
 \centering
 \subfigure[]{
   \includegraphics[angle=90, width=0.5\textwidth]{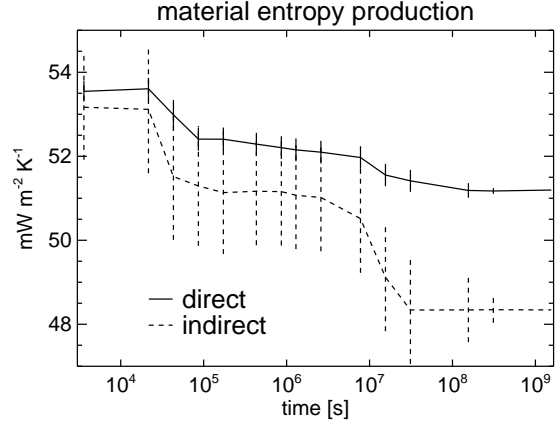}
    \label{sub1}}
   \subfigure[]{   \includegraphics[angle=90, width=0.5\textwidth]{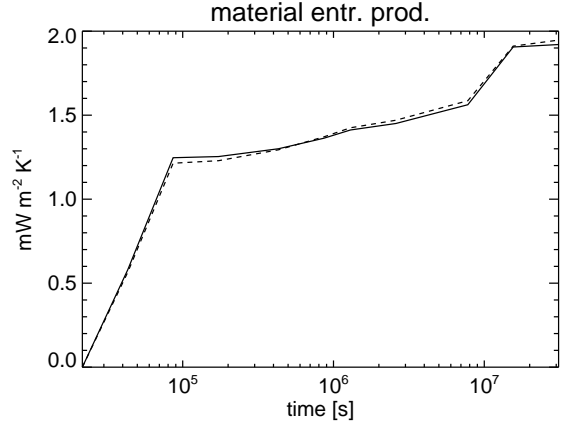}
    \label{cov1}}
   \subfigure[]{
     \includegraphics[angle=90, width=0.5\textwidth]{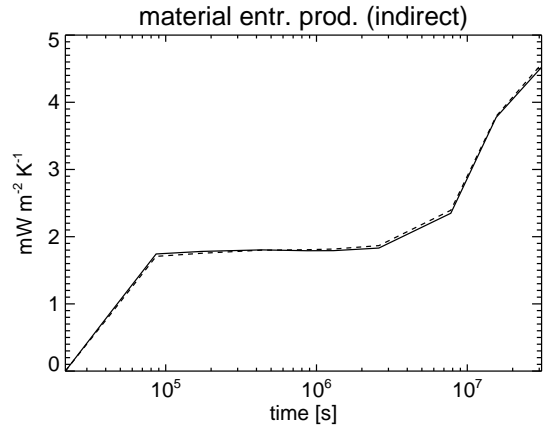}
     \label{cov2}}
\caption{ (a) Time-scale dependence of the total material entropy production (direct and indirect estimates); $\tau$ ranges between $3600$ s (model timestep) to $1.5\times 10^9$ s (50 years); (b) Differences between the exact and timeocrase grained material entropy production $\Delta\left[\overline{\dot{S}^{dir}_{mat}} \right]^{\tau}$ for $3600$ s $\le\tau\le 1$ year (c);  as in (b) but for $\Delta\left[\overline{\dot{S}^{ind}_{mat}} \right]^{\tau_{M}}$. The dashed lines represent  the correlation terms given in Eqs (14)-(15).}
  \end{figure}

In Fig.\ref{sub1} we report the estimates of the material entropy production obtained through the direct formula $ \overline{\langle \dot{S}_{mat}^{dir}\rangle^\tau}$ and the indirect formula $ \overline{\langle \dot{S}_{ind}^{dir}\rangle^\tau}$ , respectively, where we have dropped the lower index $v$ because we do not perform any spatial coarse graining. The vertical bars indicate the uncertainty due to the long term  variability. 

The computed values (worked out at each timestep) of $\overline{\dot{S}_{mat}^{ind}}\approx 53.1$ mW m$^{-2}$ K$^{-1}$ ($1mW=10^{-3}W$) and $\overline{\dot{S}_{mat}^{dir}}\approx 53.5$ mW m$^{-2}$ K$^{-1}$ have a difference of about  $0.4$ mW m$^{-2}$ K$^{-1}$, so that  Eq. (\ref{balance}) is verified with great accuracy.  The discrepancy term between the two estimates is due to the extremely small spurious radiative imbalance at TOA of about $0.1$ $W$ $m^{-2}$ (see \citep{Luc10}) and to numerical inaccuracies. Moreover, as discussed in \citep{Luc10,Pascale2}, these estimates are in good agreement with what found in climate models of higher degree of complexity. 

As conjectured, we find that the estimates of the coarse grained entropy production decrease with increasing $\tau$ from these reference values obtained with no temporal coarse graining. The bias resulting from the use of the indirect formula is larger for all values of $\tau$. In Fig.~\ref{sub1} we see that  if we consider values of $\tau$ up to $6$ hours, the impact of coarse graining is extremely small. This implies that such small time scales the irreversible processes are negligible; this matches well with the fact that convection, which is the dominating fast process in the climate system, is parametrized with an instantaneous adjustment. This immediately points to an unwelcome spurious effects of climate parametrizations. 

\begin{figure}
 \centering
\subfigure[]{
    \includegraphics[angle=90, width=0.6\textwidth]{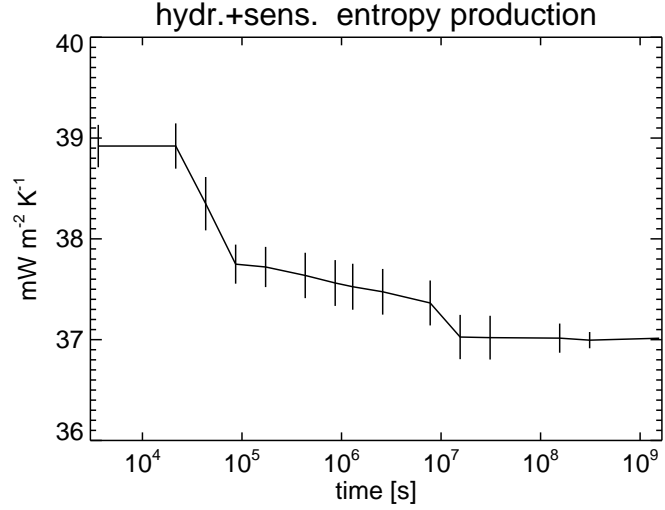}
    \label{sub2}}
       \subfigure[]{
    \includegraphics[angle=90, width=0.6\textwidth]{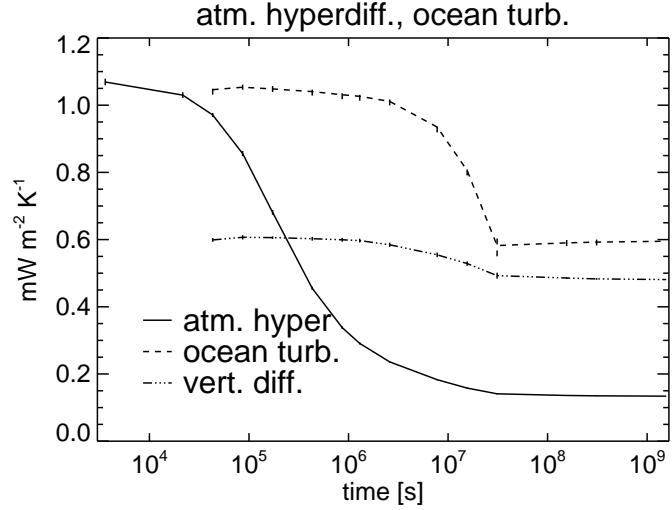}
         \label{sub3}} 
\caption{(a) Material entropy production due to hydrological cycle and heat diffusion. Here $L=50$ years. (b) Material entropy production due to atmospheric small-scale temperature diffusion (continuous line). The material entropy production due to ocean turbulence (vertical and horizontal diffusion, mixed layer physics and convection is also reported, see \cite{Pascale} for details).  We also show the vertical diffusion contribution (dotted-dashed line) to the total turbulent material entropy production. Note that the oceanic processes have a $12$-hour timestep and have not been considered in the total entropy budget .\label{cov} }
\end{figure}

The effect of coarse graining becomes more relevant when $\tau\geq1$ day. Figures ~\ref{cov1}  and \ref{cov2} present the values of  $\Delta\left[\overline{\dot{S}^{dir}_{mat}} \right]^\tau$ and $\Delta\left[\overline{\dot{S}^{ind}_{mat}} \right]^\tau$ as a function of $\tau$. For $\tau \sim 1$ day, the difference between the true and the coarse grained value of the entropy production is about $\approx 0.6$ mW m$^{-2}$ K$^{-1}$ if we use the direct formula, and $\approx 2$ mW m$^{-2}$ K$^{-1}$ is we use the indirect formula. Such biases are due to neglecting the mixing occurring on the time scale of the day, mostly due related to the daily cycle of incoming radiation. When considering the direct formula, it is interesting to note that $\Delta\left[\overline{\dot{S}^{dir}_{KE}} \right]^\tau$ is basically zero for all values of $\tau$ (not shown), meaning that there is no time correlation between the dissipation of kinetic energy and the  temperature field. %The dissipation of kinetic energy is usually small, and does not influence strongly the local temperature; nonetheless, it plays an important role because, as discussed above, it is a positive definite contribution to the entropy production which \text{resists} all coarse graining procedures.  
The coarse graining, instead, impacts the contribution to entropy production due to the hydrological cycle. We can substantiate this statement by observing that  $\Delta\left[\overline{\dot{S}^{dir}_{mat}} \right]^\tau \sim \Delta\left[\overline{\dot{S}^{dir}_{heat}} \right]^\tau$ (see definition of the latter in Eq. \ref{matter}), as can be seen by comparing Figs. \ref{cov2} and ~\ref{sub2}. 

The second timescale worth discussing is the one corresponding to $1$ year ($\sim 3\times 10^7$ s). The use of annual means instead of time-step data  introduces a   bias of about $4$ mW  m$^{-2}$ K$^{-1}$ when using the indirect formula, which corresponds to neglecting the correlation between the seasonal cycle of the radiation budget and that of the radiation temperature field. Similarly, considering the direct formula, we obtain $\Delta\left[\overline{\dot{S}^{dir}_{mat}} \right]^\tau \sim 1.5$ mW  m$^{-2}$ K$^{-1}$, which measures  the effect of neglecting the correlation of the seasonal cycle of the atmospheric and oceanic transport and dissipation and of the  temperature field. One must note that a considerable contribution to the value of $\Delta\left[\overline{\dot{S}^{dir}_{mat}} \right]^\tau$ for $\tau\geq 1$ year is given by the atmospheric temperature hyperdiffusion,
which in FAMOUS is implemented as a eight-order laplacian operator and applied after the advection to the model prognostic variables. Hyperdiffusion is generally introduced in dynamic cores for numerical reasons  in order to smooth variables and avoid local divergences.  However it may thought  as a way to represent   turbulent dissipation and mixing  at subgrid scale. We discover that a traditional numerical \textit{trick} used in the climate modeling community for avoiding computational instabilities impacts a global scale physical properties of the system, as observed in \cite{LucariniRagone} when looking at energy budgets.

We also observe that there is no clear signature emerging in the functions  $\Delta\left[\overline{\dot{S}^{dir}_{mat}} \right]^\tau$ and $\Delta\left[\overline{\dot{S}^{ind}_{mat}} \right]^\tau$ for values of $\tau$ to weeks (synoptic waves) or monthly (low frequency variability) time scales while a relatively smooth transitions is found going from daily to yearly averages. This supports the idea that it is possible to look at weather disturbances as parts of a macro-turbulent cascade. 

%Therefore, our analysis quantifies the impact of the two main time modulations of the solar forcing on the climate system on the entropy production.

Estimating the entropy production via either the direct or the indirect formula using long term averages (but full spatial resolution) leads to underestimate the exact value of the entropy production by less than 10\%. This suggests that long term averages of the climatic fields one can obtain from the climate repositories are enough to get a good idea of the properties of the climate system. As we shall see in the next section, things change drastically when the coarse graining impacts the spatial features of the climatic fields. 

We conclude this section with a note on the oceanic processes, which we do not treat in this paper as they contribute negligibly to the overall entropy production in the climate system. in Fig. \ref{sub2} we show the dependence of the entropy production due to the oceanic mixing on the temporal coarse graining (dashed) line. We discover that its exact value, computed at time step, is about $1$ mW  m$^{-2}$ K$^{-1}$, as in \citep{Pascale2}, and its coarse grained value does not noticeably decrease up to $\tau\sim 1$ year, above which the coarse grained estimate is roughly halved. The dash-dotted line in Fig.  \ref{sub2} gives the contribution due to the vertical mixing in the interior of the ocean, which is a very slow process and is, in fact, weakly affected by the temporal coarse graining.  The other contribution to the entropy production in the ocean comes from the mixing occurring in the mixed layer. The mixing layer scheme \cite{Kraus} parametrizes the convection due to heating at depth and cooling at the surface as well as the mechanical stirring due to wind and is introduced in ocean models in order to account for the seasonal thermocline variations. The coarse grained value of this term goes virtually to zero for $\tau\geq 1$ because, when considering such an averaging, we discount for the impact of the seasonal cycle in the upper portion of the ocean. 

\subsection{Spatial and Temporal coarse-graining}\label{spacetime}

In this section we analyze the combined effect of coarse graining the  heating rates and temperature fields  in space and time by using extensively Eqs. \ref{cgrainingdir} and \ref{cgrainingind}. % We remind that at full resolution the spatial discretization of the atmospheric variables is described by a regular grid with angle resolution of $5^{\circ}\times 7.5^{\circ}$ lat-lon and by 11 vertical levels is $\sigma$ coordinates. 
Of course, there are many ways to perform coarse graining, boiling down to the selection of the stencil $v$ introduced before. Summarizing, we proceed as follows:
\begin{enumerate}
\item longitudinal averaging: the stencils $v$ are given by arcs of varying length in the zonal direction;  
%\item latitudinal averaging: the stencils $v$ are given by arcs  of varying length in the longitudinal direction;
\item areal averaging: the stencils $v$ are given by portions of varying size of the spherical surface;
\item mass averaging: the stencils $v$ are given by same-mass portions of the atmospheric spherical shell obtained by thickening in the vertical direction the stencils described in 2.;
\end{enumerate}
in all cases we perform also temporal coarse graining by selecting the same averaging times $\tau$ described in the previous subsection. 

%We underline that, given the resolution of the model output, choosing stencils $v$ with the same mass throughout the volume of the climate system would require heavy remapping on the thermodynamic fields, which would lead to spurious contributions to the entropy production. Instead, 

It is important to note that the averaging as in points 1. and 2 is performed at constant $x_3$. FAMOUS (and HadCM3) uses  hybrid vertical coordinates, \textit{i.e.} a coordinate system which changes smoothly from a terrain-following specification near the lower boundary ($\sigma$ coords.) to a isobaric definition ($p$ coords.) in the medium-upper troposphere and stratosphere. As clear from Eqs. \ref{cgraining2}-\ref{cgrainingind}, the result of any coarse graining performed at constant value of the vertical coordinates depends on the vertical coordinate considered.  In order to avoid the spurious effects of remapping the thermodynamic fields to a new coordinate system, we choose coarse grained grid boxes that respect as much as possible the original model's resolution.

We also remark that given the heavy computational burden of the operation, we restrict our analysis to only one of the fifty year of available data. We have tested that 
%Details are given in Tab.~\ref{tabellina}.
%Here we study the effect of combined space and time averaging. We start from the full model resolution (gaussian grids with angle resolution of $7.5^{\circ}\times 5^{\circ}$ lon-lat), which we denote as $\mathbf{x}_{full}$. Then we define a new grid with a coarser resolution, say $\mathbf{x}_r$ and convert  the fields  $\overline{q(\mathbf{x}_{full}, t_j)}^{\tau}$ to $\overline{q(\mathbf{x}_r, t_j)}^{\tau}$ and $\overline{T(\mathbf{x}_{full}, t_j)}^{\tau}$ to $\overline{T(\mathbf{x}_r, t_j)}^{\tau}$.  The values on the new grid is obtained as an area-average of the parts of the boxes  on the old grid which overlap with the box on the new grid.  We will gradually  decrease the space resolution by interpolating  the heating rates and temperature fields over coarser and coarser  grid domains. 

\subsubsection{Longitudinal Averaging}

\begin{figure}
 \centering
 \subfigure[]{
    \includegraphics[angle=90, width=0.46\textwidth]{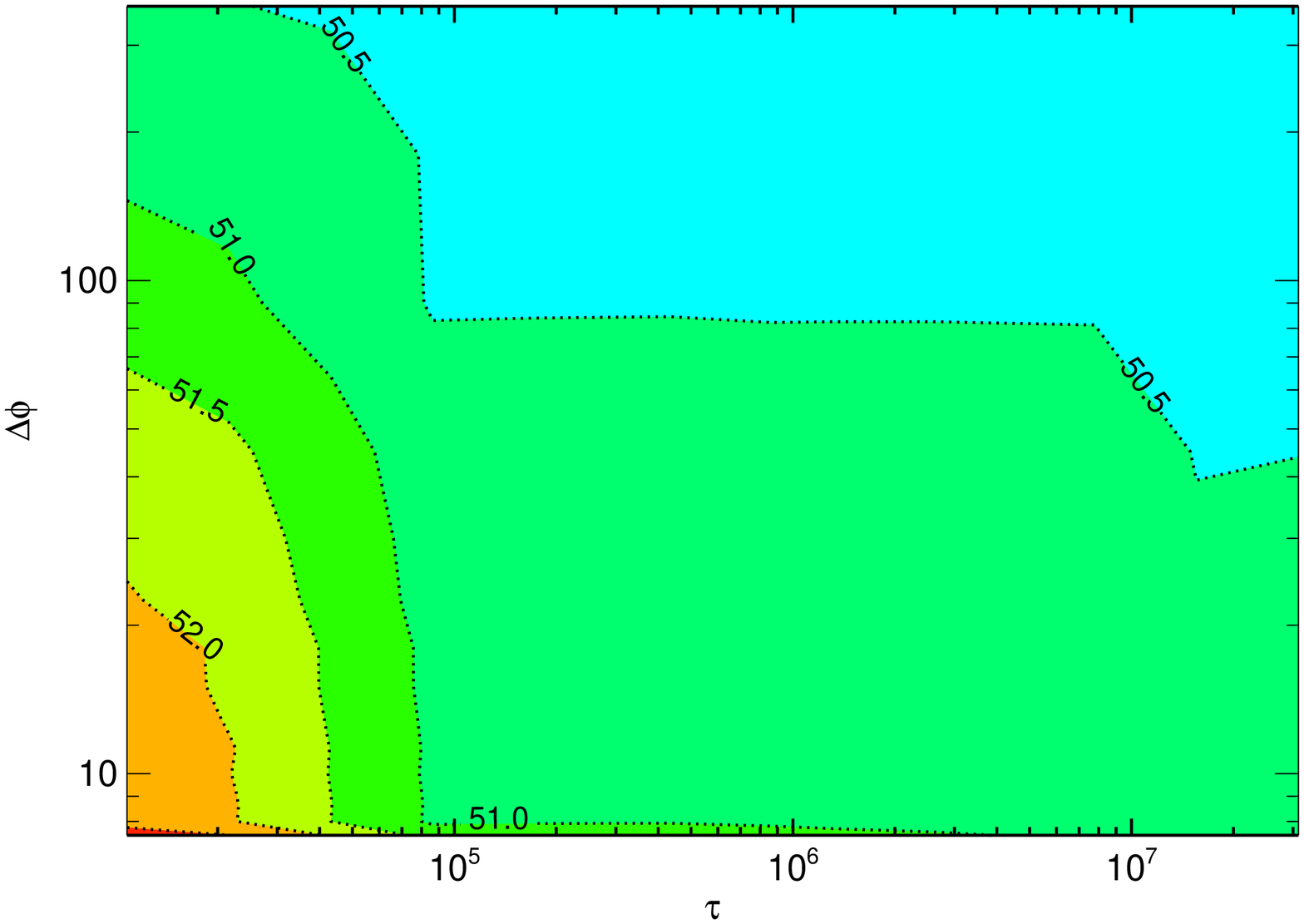}
    \label{zon1}
   }
 \subfigure[]{
    \includegraphics[angle=90, width=0.46\textwidth]{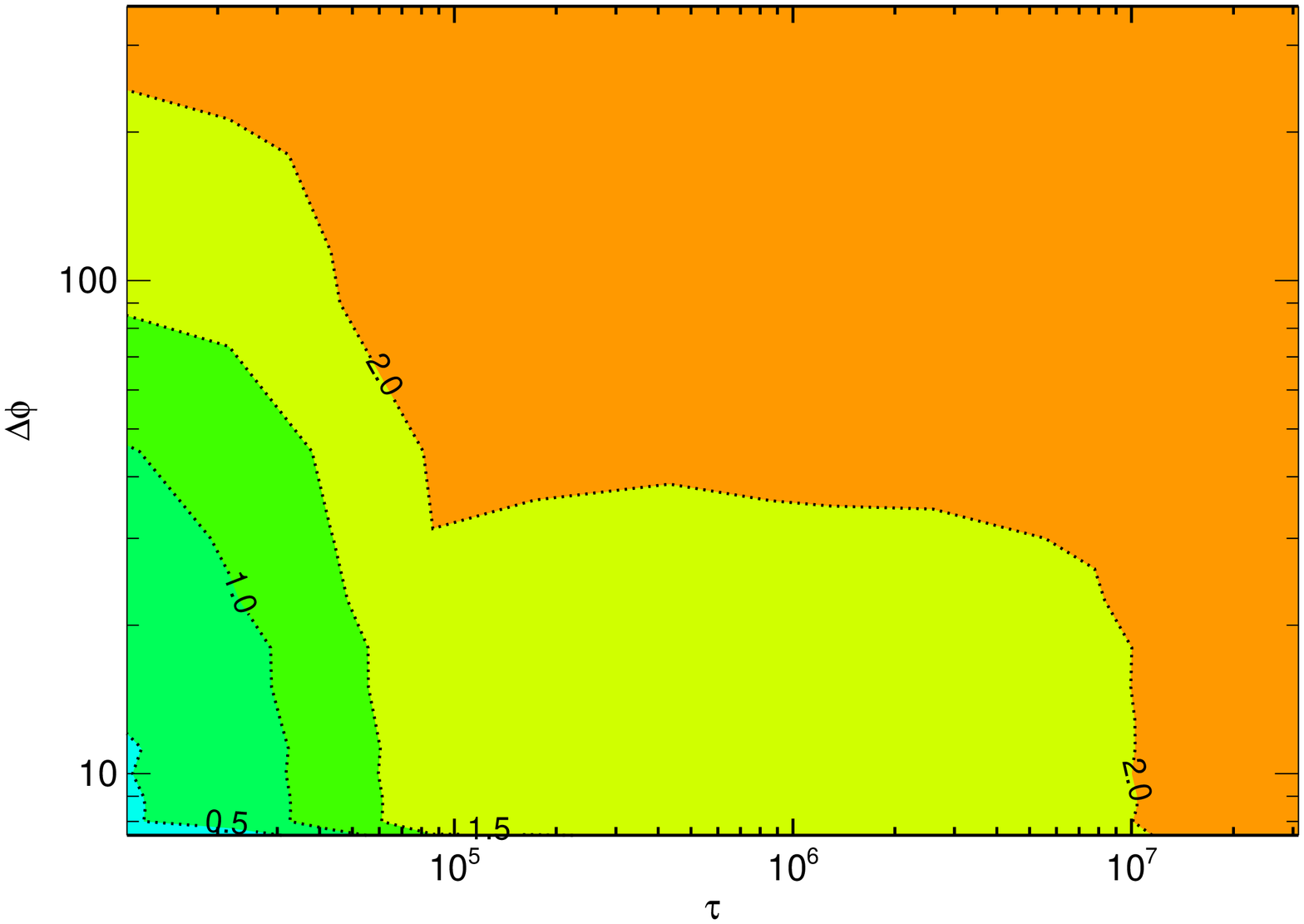}
    \label{zon1b}
   }  
   \subfigure[]{
     \includegraphics[angle=90, width=0.46\textwidth]{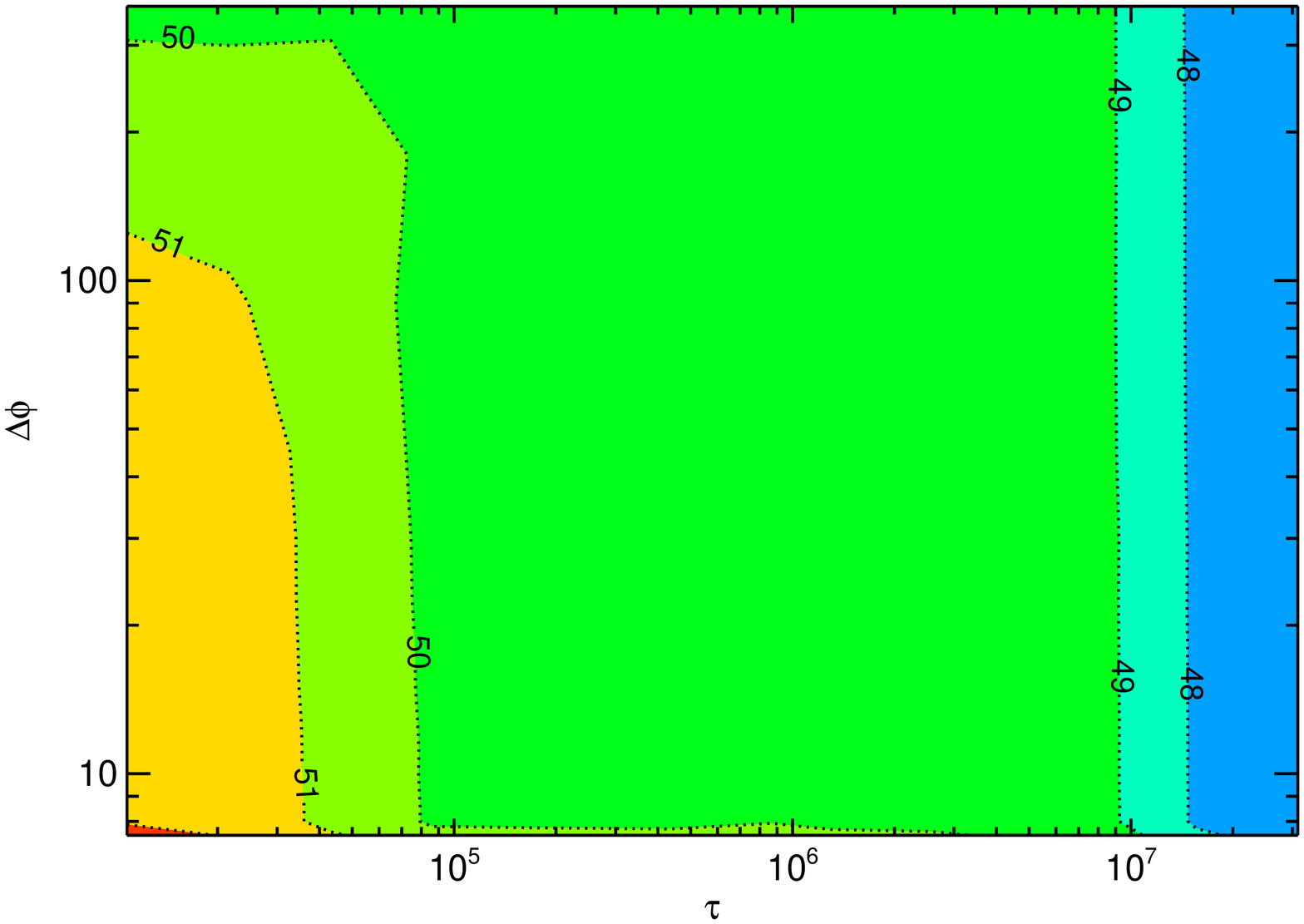}
     \label{zon2}
      } 
 \subfigure[]{
    \includegraphics[angle=90, width=0.46\textwidth]{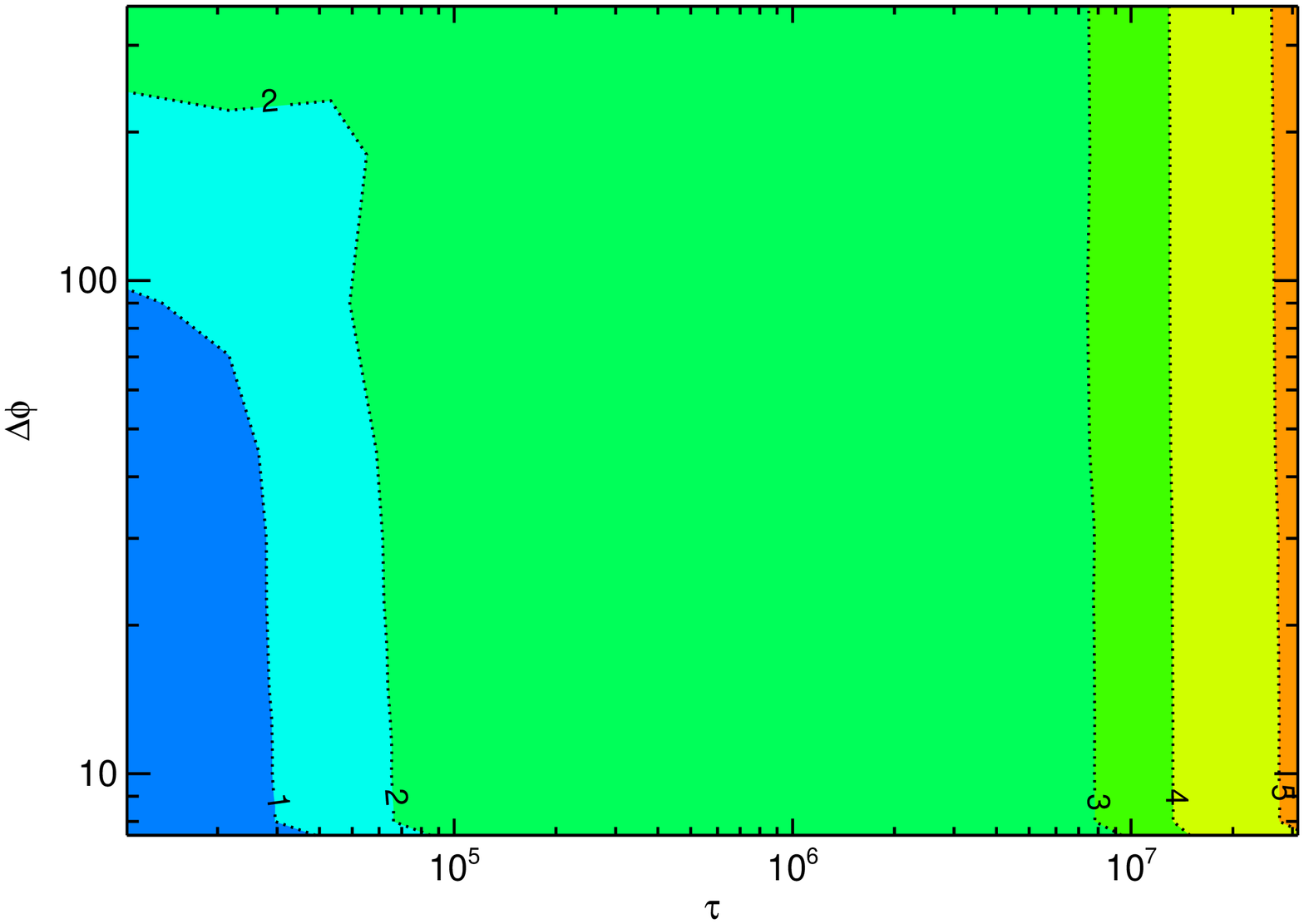}
    \label{zon2b}}
\caption{ Estimates of $\overline{\langle\dot{S}^{did}_{mat}\rangle_v^\tau}$ (a), $\Delta\left[\overline{\dot{S}^{dir}_{mat}} \right]_v^\tau$ (b), $\overline{\langle\dot{S}^{ind}_{mat}\rangle_v^\tau}$ (c), and $\Delta\left[\overline{\dot{S}^{ind}_{mat}} \right]_v^\tau$ (d). The spatial averaging considered here is given by longitudinal averages on horizontal surface. The $x$-axis reports $\tau$, the $y$-axis describes the extent $\Delta \phi$ of the averaging in $^{\circ}$. \label{zon}  }
\end{figure}

We first investigate the effect of coarse graining on the estimate of the material entropy production by averaging longitudinally the thermodynamic fields, up to the point of considering zonally averaged only fields, and by degrading their temporal resolution by using the averaging times $\tau$ described above. The estimates of $\overline{\langle\dot{S}^{dir}_{mat}\rangle_v^\tau}$ and $\overline{\langle\dot{S}^{ind}_{mat}\rangle_v^\tau}$ are given in  Fig.~\ref{zon1} and  Fig.~\ref{zon2}, respectively. The corresponding values of $\Delta\left[\overline{\dot{S}^{dir}_{mat}} \right]_{v}^\tau $ and $\Delta\left[\overline{\dot{S}^{ind}_{mat}} \right]_{v}^\tau$ are reported in Fig. \ref{zon1b} and  Fig.~\ref{zon2b}, respectively, and some specific results are given in Table \ref{tabellina2}. 

In all figures, the value of $\tau$ is reported in the abscissae, while in the ordinates  the value of size of the spatial stencil, ranging from $7.5^\circ$ (no coarse graining) to $360^\circ$ (zonal averaging) is shown. In both figures, the lower left corner corresponds to the best estimate of the entropy production; the upper left corner corresponds to the material entropy production due to the longitudinally averaged, high-temporal resolution fields. the lower right corner corresponds to long-time averaged, high-resolution spatial case, and, eventually, the upper right corner corresponds to the highest degree of coarse graining: it represent the entropy production due to the long-term averaged, longitudinally averaged fields, and features the lowest value of  $\overline{\langle\dot{S}^{dir}_{mat}\rangle_v^\tau}$ and $\overline{\langle\dot{S}^{ind}_{mat}\rangle_v^\tau}$. We remark that the values reported at the border of the domain given by the lowest value of the ordinates coincide, obviously, with what shown in Fig. \ref{sub1}. As a general fact, we observe that the coarse grained estimates of the entropy production decrease (or remain virtually unchanged) as we perform coarser and coarser graining procedure, in time or in space, and that $\overline{\langle\dot{S}^{dir}_{mat}\rangle_v^\tau}\geq\overline{\langle\dot{S}^{ind}_{mat}\rangle_v^\tau}$.

The strongest dependence of the $\overline{\langle\dot{S}^{ind}_{mat}\rangle_v^\tau}$  is on $\tau$: temporal coarse graining appears to be the dominating influence, while the effect of spatial coarse graining is apparent only for $\tau\leq1$ day and for considerable longitudinal averaging, such that features below $60^\circ$ are smeared out. In other terms, longitudinal averaging starts to matter only when we lose information on the alternating pattern continents/oceans. The total effect of removing totally the spatial structure is similar to that of performing a time-averaging of one day. When considering $\overline{\langle\dot{S}^{ind}_{mat}\rangle_v^\tau}$ the picture is partially different: first, the influence of the spatial averaging is relatively strong at all scales for $\tau\leq1$ day. The coupling between the spatial and temporal scales indicates that using low pass filter and space and time we remove the fast traveling synoptic waves of the mid-latitudes. As opposed to the case of the coarse grained indirect estimate of the entropy production, spatial averaging plays a role also for $1$ day $\leq\tau\leq3$ months.  This is probably the signature of the relevance of low-frequency, large scale features of the tropical circulation, which are sustained by longitudinal gradients (and tend to reduce them), which are smeared out when extreme coarse graining  is applied.   

Concluding, we remark that neglecting information on the longitudinal fluctuations and temporal fluctuations of the thermodynamic fields does not bias considerably (in the worst case, by about 10\%) the estimates of the entropy production one would obtain by retaining the full information. This agrees with the fact that, in first approximation, in our planet longitudinal gradients and longitudinal heat fluxes are relatively small \cite{Peix2}. Moreover, using either the direct or indirect formula we obtain rather similar results, with a bias of maximum 5\%. As expected, the direct formula gives more accurate estimates for all considered coarse graining procedures.   

%sharp drop at most of timescale after first coursing step ($n_{\varphi}$ from 48$^{\circ}$ to 45$^{\circ}$);  general feature $\dot{S}_{mat}$ decreases when both $\tau$ and $\Delta\varphi$ is increased. In particular decreasing  the zonal resolution is equivalent to getting rid of zonal gradients. The entropy production we loose is therefore the bit associated with the zonal energy flows (breeze and due mostly to land-ocean contrasts). From both $\dot{S}_{mat}^{dir}$  and $\dot{S}_{mat}^{ind}$ at the highest time resolution (timestep) this is nearly  $2$ mW m$^{-2}$ K$^{-1}$. 

%The dependance on the spatial resolution becomes weaker and weaker at higher time averaging periods.

\subsubsection{Areal averaging}
%\label{spaceres}

As a second step for understanding the role of spatial and temporal coarse graining of the estimate of the material entropy production, we combine time averaging of the thermodynamic fields with areal averaging along horizontal surfaces. This operation allows us to explore how the two-dimensional spatial covariance of heating and temperature fields contributes to entropy production at all time scales. In order to keep coherence with the previous coarse graining procedure, we proceed as follows. We divide the spherical surface in coarse grained grid boxes defined by intervals (in degrees in latitude and longitude) $(\Delta \lambda\Delta \phi)$ such that $\Delta \phi/\Delta \lambda=1.5$, which is consistent with the model's resolution of $5^\circ lat\times 7.5^\circ lon$. We then increase $\Delta \phi$ from $7.5^\circ$ up to $90^\circ$, thus decreasing progressively the number of coarse grained grids from 1728 to 12. In order to complete the coarse graining, we select as  two coarsest resolutions $(\Delta \lambda;\Delta \phi)=(90^\circ,180^\circ)$ (four quadrants) and $(\Delta \lambda;\Delta \phi)=(180^\circ,360^\circ)$ (full spherical surface).

\begin{figure}
\centering
 \subfigure[]{
    \includegraphics[angle=90, width=0.46\textwidth]{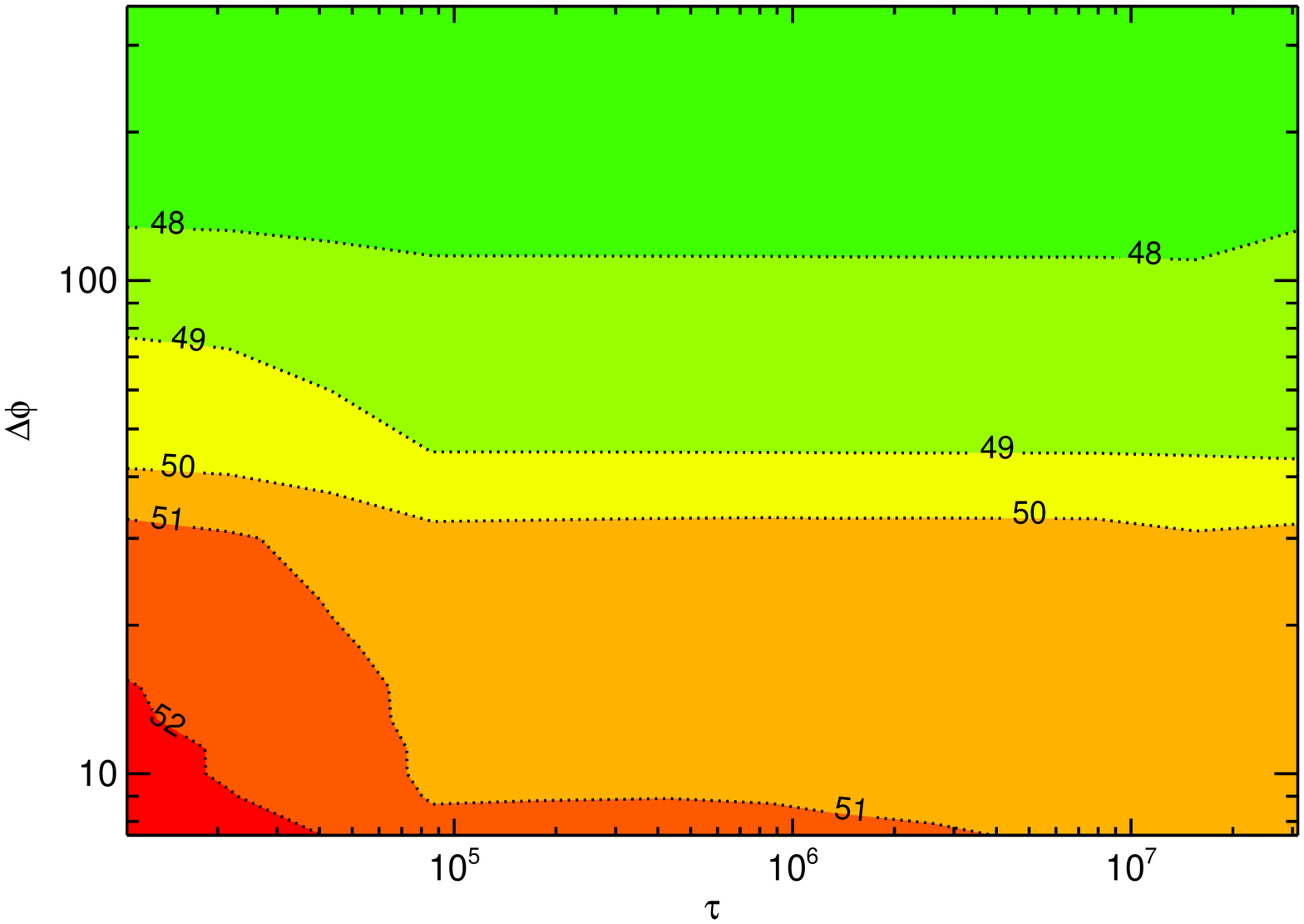}
    \label{space1}
   }
 \subfigure[]{
    \includegraphics[angle=90, width=0.46\textwidth]{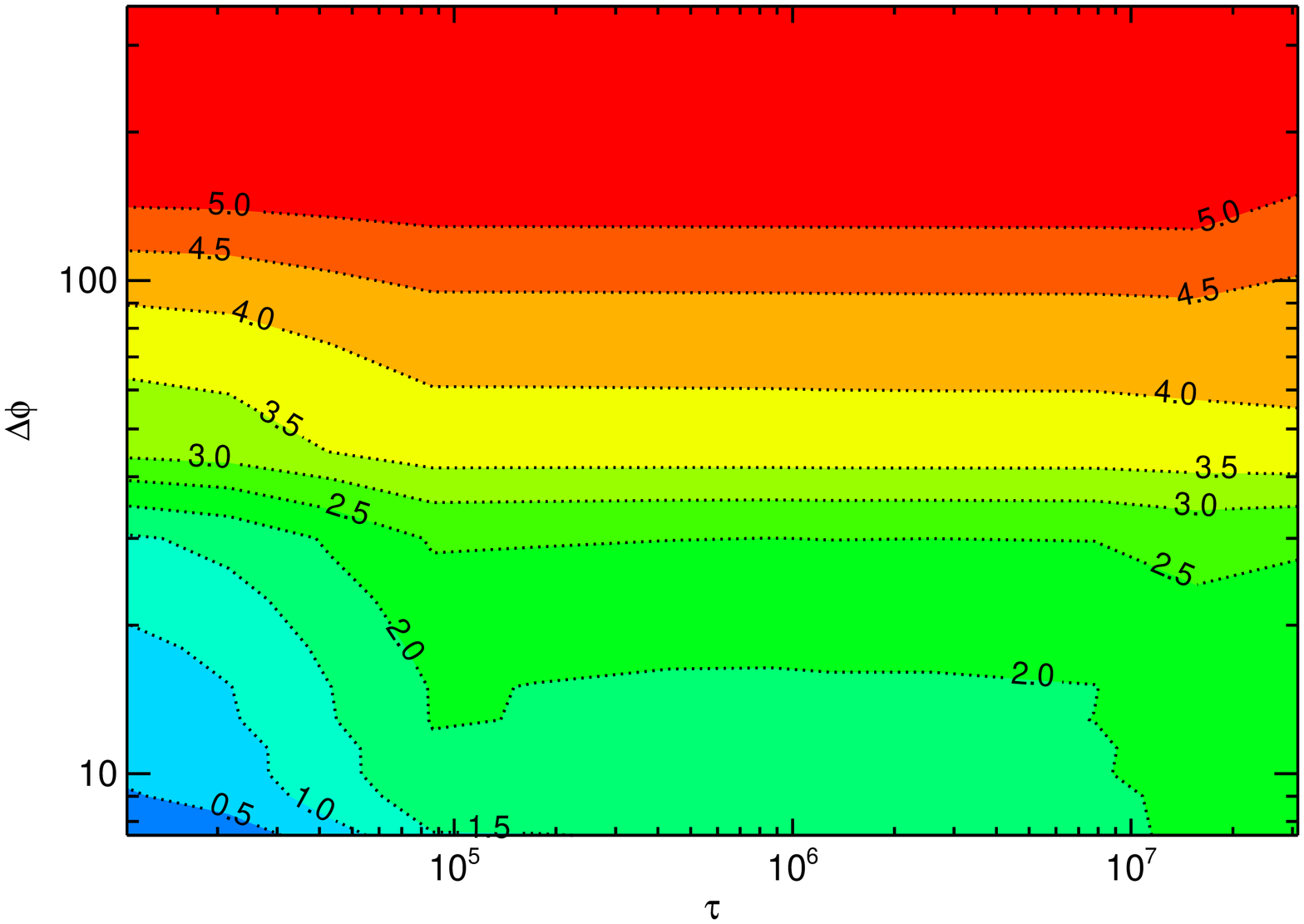}
    \label{space1b}
   }
   \subfigure[]{
     \includegraphics[angle=90, width=0.46\textwidth]{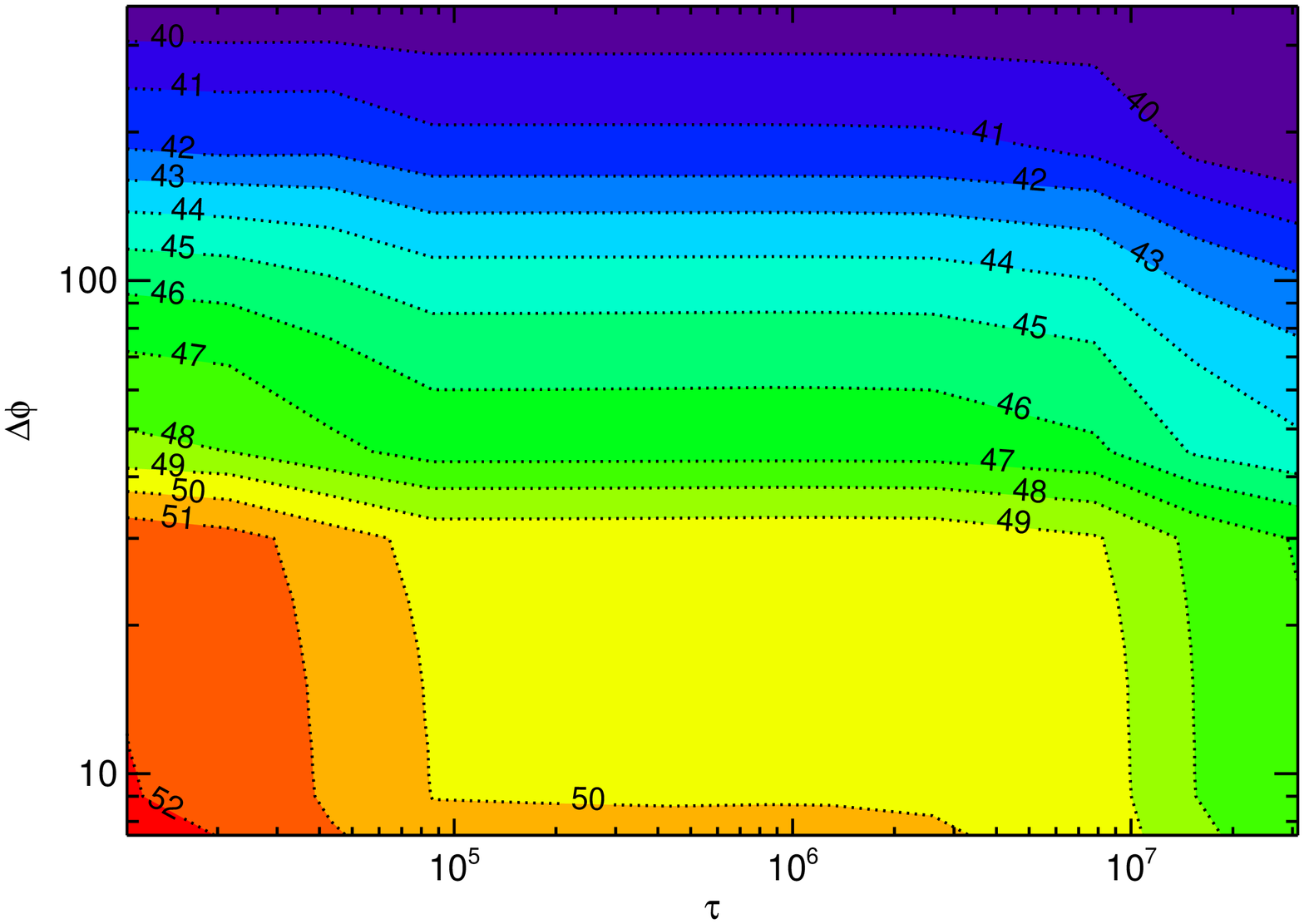}
     \label{space2}
      } 
 \subfigure[]{
    \includegraphics[angle=90, width=0.46\textwidth]{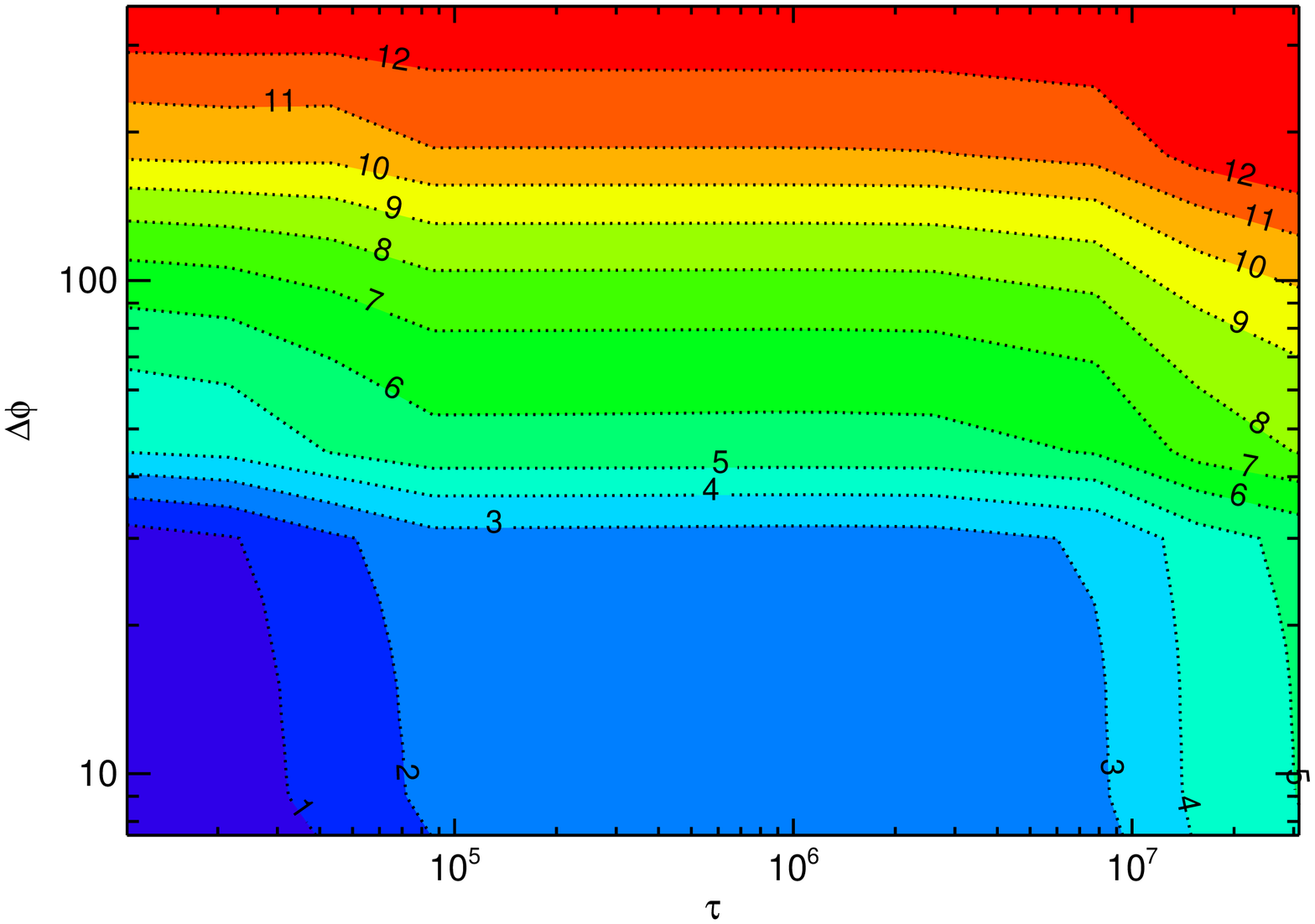}
    \label{space2b}
   }
\caption{Estimates of $\overline{\langle\dot{S}^{did}_{mat}\rangle_v^\tau}$ (a), $\Delta\left[\overline{\dot{S}^{dir}_{mat}} \right]_v^\tau$ (b), $\overline{\langle\dot{S}^{ind}_{mat}\rangle_v^\tau}$ (c), and $\Delta\left[\overline{\dot{S}^{ind}_{mat}} \right]_v^\tau$ (d). The spatial averaging considered here is given by areal averages along horizontal surfaces. The $x$-axis reports $\tau$, the $y$-axis describes the longitudinal extent $\Delta \phi$ of the coarse grained grid boxes. Details are given in the text. \label{equi} }
\end{figure}

 The estimates of $\overline{\langle\dot{S}^{dir}_{mat}\rangle_v^\tau}$ and $\overline{\langle\dot{S}^{ind}_{mat}\rangle_v^\tau}$ are given in  Fig.~\ref{space1} and  Fig.~\ref{space2}, respectively. The corresponding values of $\Delta\left[\overline{\dot{S}^{dir}_{mat}} \right]_{v}^\tau $ and $\Delta\left[\overline{\dot{S}^{ind}_{mat}} \right]_{v}^\tau$ are reported in Fig. \ref{space1b} and  Fig.~\ref{space2b}, respectively, and some specific results are given in Table \ref{tabellina2}. We discover that, as opposed to the previous case, the impact of selecting coarser and coarser graining in space reduces considerably the value of  $\overline{\langle\dot{S}^{dir}_{mat}\rangle_v^\tau}$ for all values of $\tau$, because such an averaging progressively removes the strong meridional dependence of the thermodynamic fields, up to the extreme case of $v$ being the whole Earth's surface. In this case, the estimate of the entropy production $\overline{\langle\dot{S}^{dir}_{mat}\rangle_v^\tau}\sim47.2$ $mWm^{-2}K^{-1}$.
 Note that when considering very strong spatial averaging the effect of changing $\tau$ is negligible, because the spatial averaging alone reduces the temporal correlations by mixing areas of the planet experiencing, e.g., different seasons.  The $\tau$-dependence of $\overline{\langle\dot{S}^{dir}_{mat}\rangle_v^\tau}$ is relevant only for $\tau\leq1$ day and spatial scales smaller than $\sim 3-4\times10^6$ $m$, which is, like in the previous case, hints at the fact that when averaging over large spatial and temporal scales, we remove the variability corresponding to synoptic waves.  

The function $\overline{\langle\dot{S}^{ind}_{mat}\rangle_v^\tau}$ has a qualitatively similar but quantitatively stronger dependence on $\Delta \phi$ and $\tau$ with respect to  $\overline{\langle\dot{S}^{did}_{mat}\rangle_v^\tau}$: the bias in the estimate production is larger for all the considered coarse graining. Additionally, the indirect formula is more strongly affected by averaging over long time scales $\tau$, similar to what seen in  Fig.~\ref{zon2}, because the coupling between the seasonal cycle of the radiative budget and the temperature fields is very strong. Note that when we consider global or quasi-global spatial coarse graining, such effect disappears, because averaging such large scales already removes a large part of the season cycle signal. This is different from what reported in Fig.~\ref{zon2}, because zonal averaging, obviously, cannot remove the asymmetry between northern and southern hemisphere. 

\subsubsection{Mass Averaging}

\begin{figure}
 \centering
% \subfigure[]{
 %   \includegraphics[angle=90, width=0.46\textwidth]{horizontal_vertical_timestep_direct.ps}
 %   \label{vhdir1}
%   }
% \subfigure[]{
 %   \includegraphics[angle=90, width=0.46\textwidth]{vera-timestep_coarse_horizontal_vertical_direct.ps}
 %   \label{vhdir1b}
%   }
   \subfigure[]{
     \includegraphics[angle=90, width=0.46\textwidth]{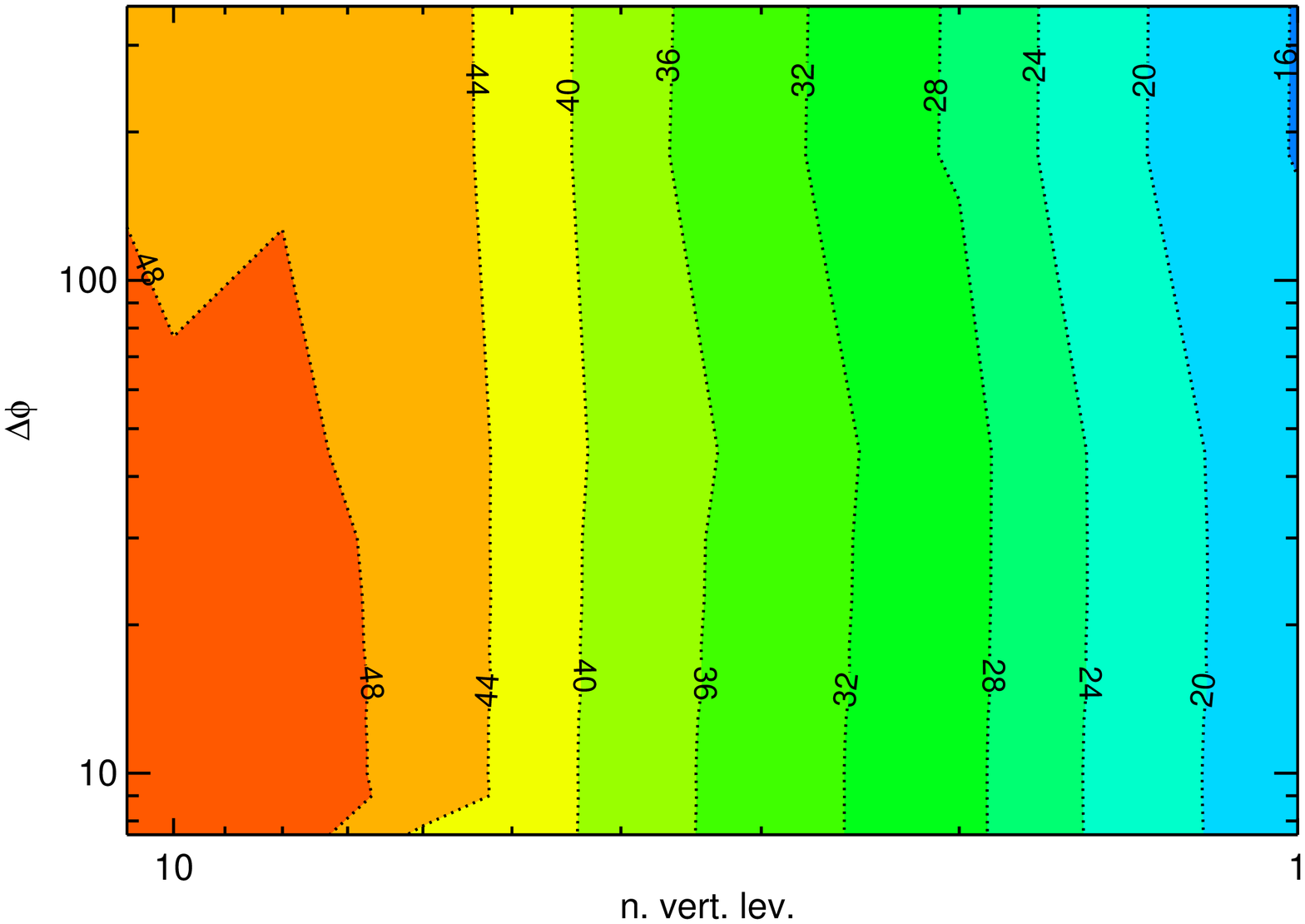}
     \label{vhdir2}
      } 
   \subfigure[]{
     \includegraphics[angle=90, width=0.46\textwidth]{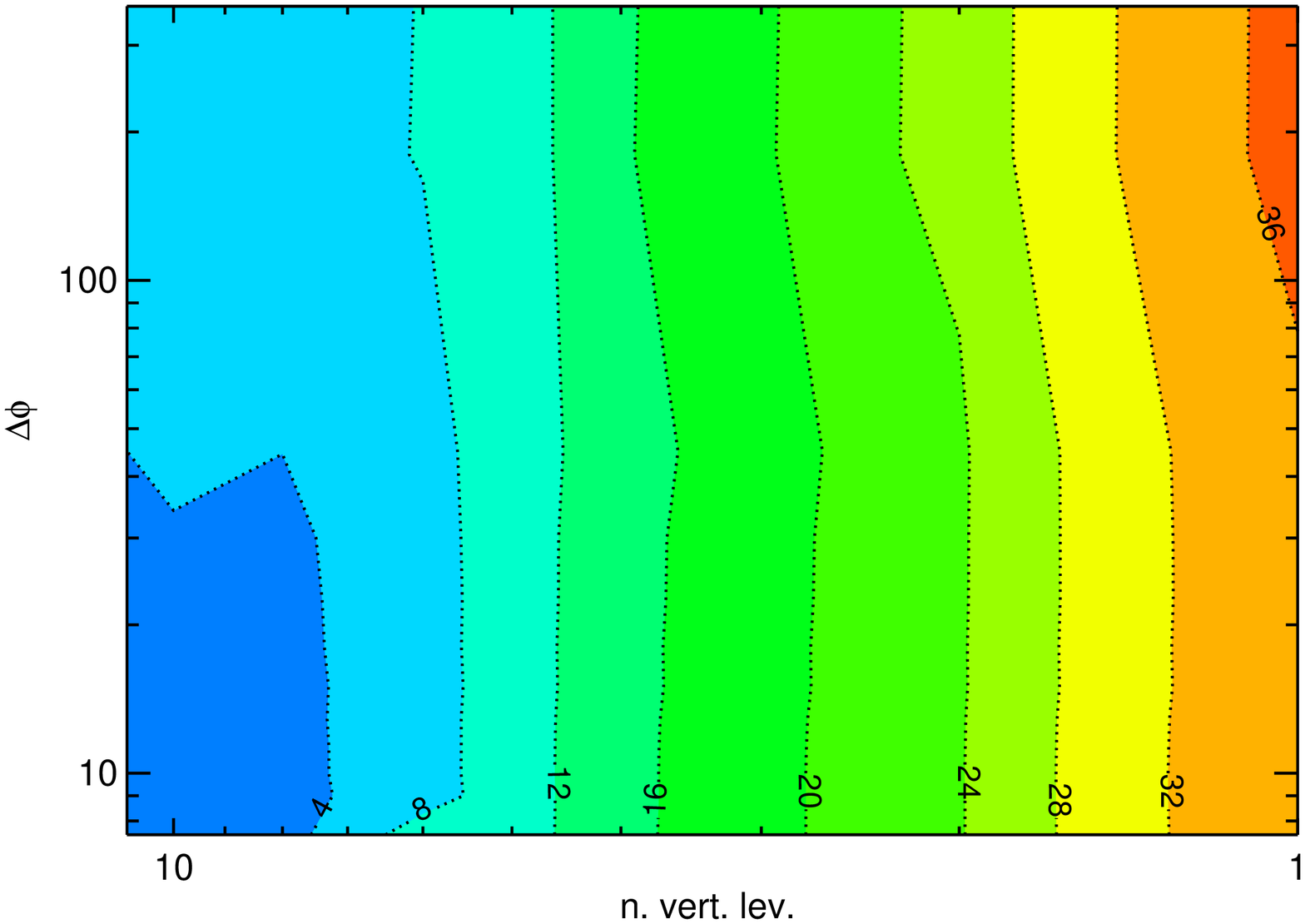}
     \label{vhdir2b}
      } 
   \subfigure[]{
     \includegraphics[angle=90, width=0.46\textwidth]{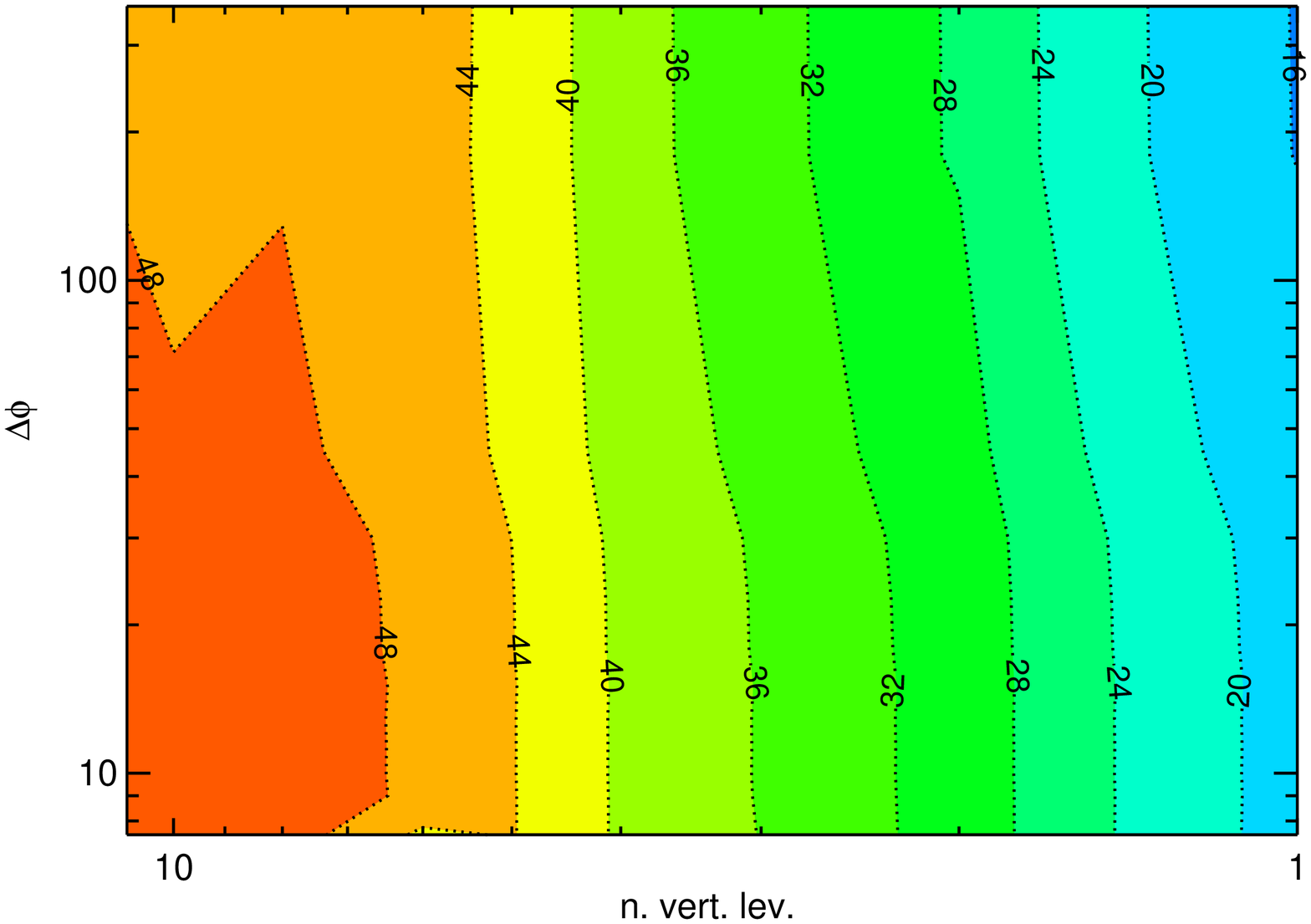}
     \label{vhdir3}
      } 
   \subfigure[]{
     \includegraphics[angle=90, width=0.46\textwidth]{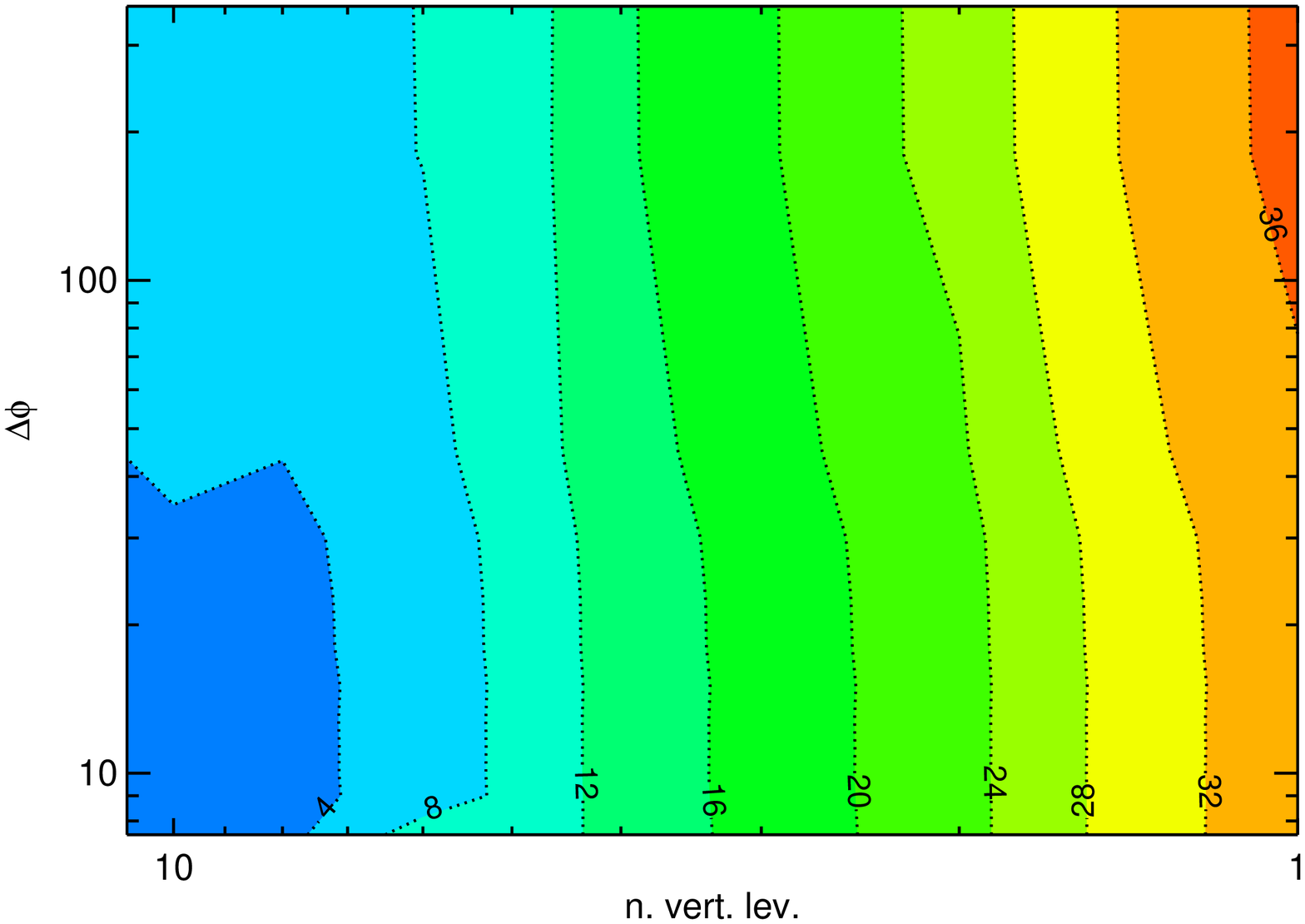}
     \label{vhdir3b}
      } 
\caption{Estimates of $\overline{\langle\dot{S}^{dir}_{mat}\rangle_v^\tau}$ (a), and $\Delta\left[\overline{\dot{S}^{dir}_{mat}} \right]_v^\tau$ (b) for $\tau=1$ day, and of  $\overline{\langle\dot{S}^{dir}_{mat}\rangle_v^\tau}$ (c), and $\Delta\left[\overline{\dot{S}^{dir}_{mat}} \right]_v^\tau$ (d) for $\tau=1$ year. The spatial averaging considered here is given by areal averages along horizontal surfaces and vertical averages along columns. The $x$-axis reports the numbers of vertical levels involved in the averaging, the $y$-axis describes the longitudinal extent $\Delta \phi$ of the coarse grained grid boxes. Details are given in the text.  \label{vhdir} }
\end{figure}

\begin{figure}
 \centering
% \subfigure[]{
 %   \includegraphics[angle=90, width=0.46\textwidth]{horizontal_vertical_timestep_indirect.ps}
 %   \label{vhind1}
%   }
% \subfigure[]{
 %   \includegraphics[angle=90, width=0.46\textwidth]{vera-timestep_coarse_horizontal_vertical_indirect.ps}
 %   \label{vhind1b}
%   }
   \subfigure[]{
     \includegraphics[angle=90, width=0.46\textwidth]{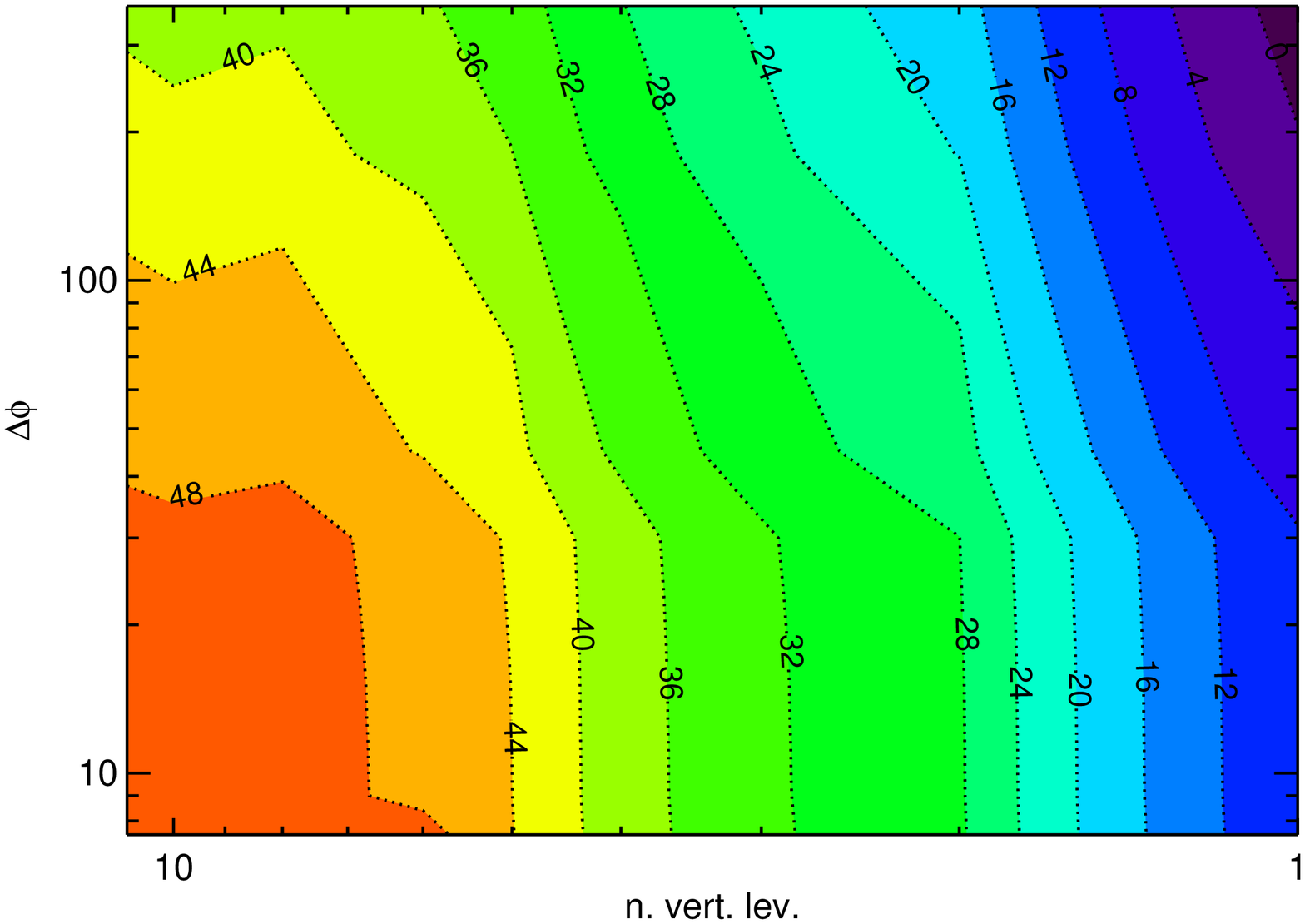}
     \label{vhind2}
      } 
   \subfigure[]{
     \includegraphics[angle=90, width=0.46\textwidth]{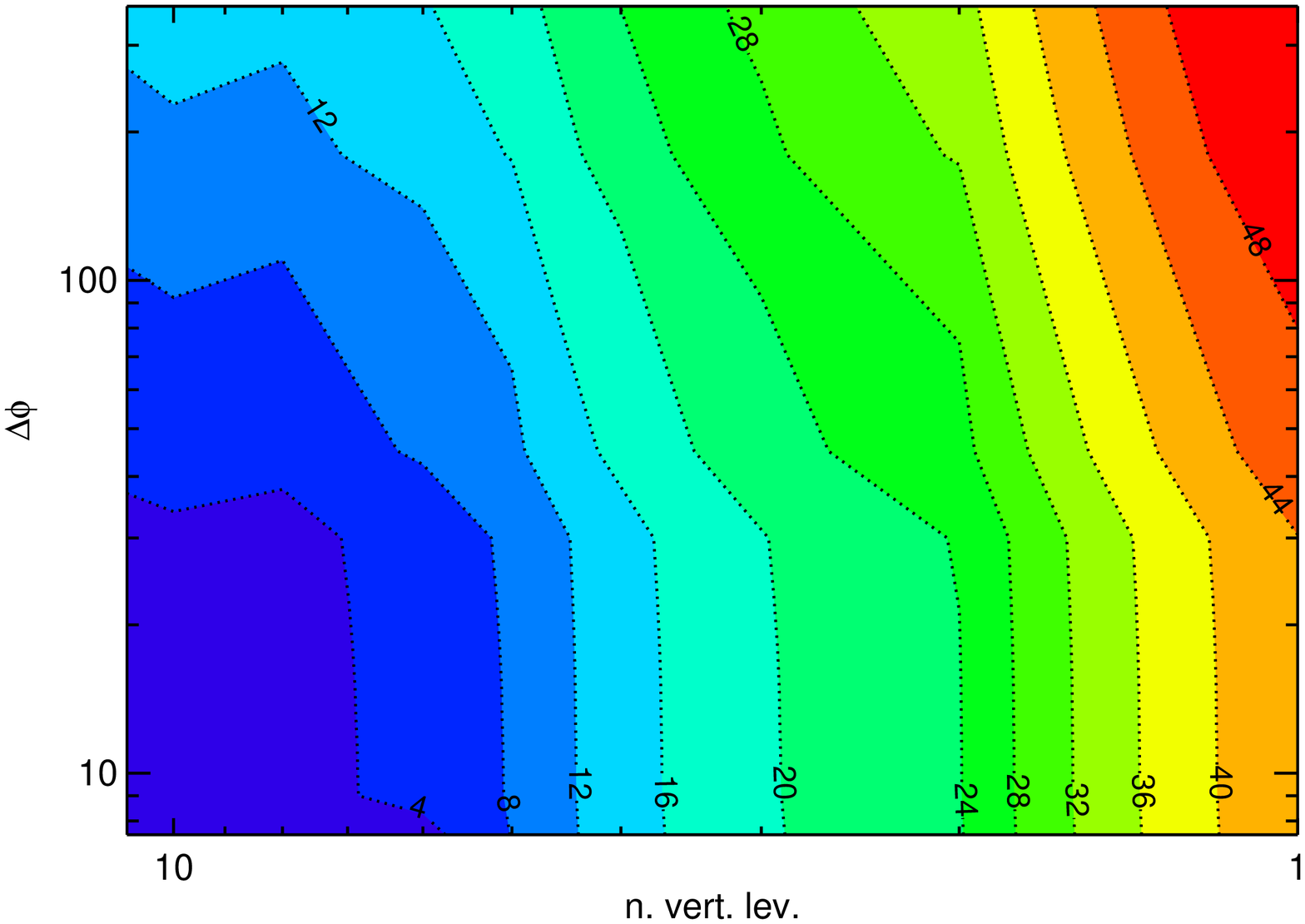}
     \label{vhind2b}
      } 
   \subfigure[]{
     \includegraphics[angle=90, width=0.46\textwidth]{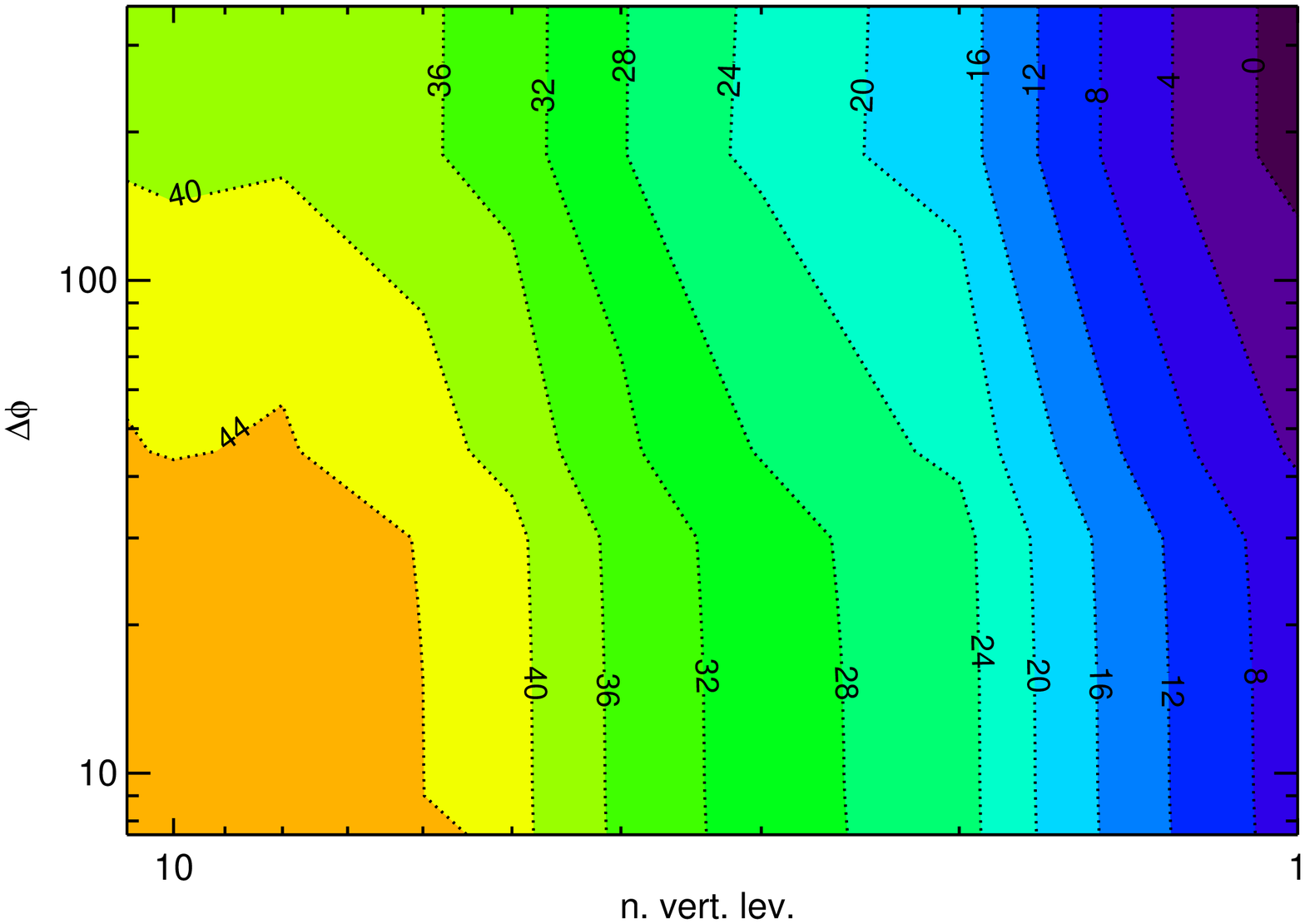}
     \label{vhind3}
      } 
   \subfigure[]{
     \includegraphics[angle=90, width=0.46\textwidth]{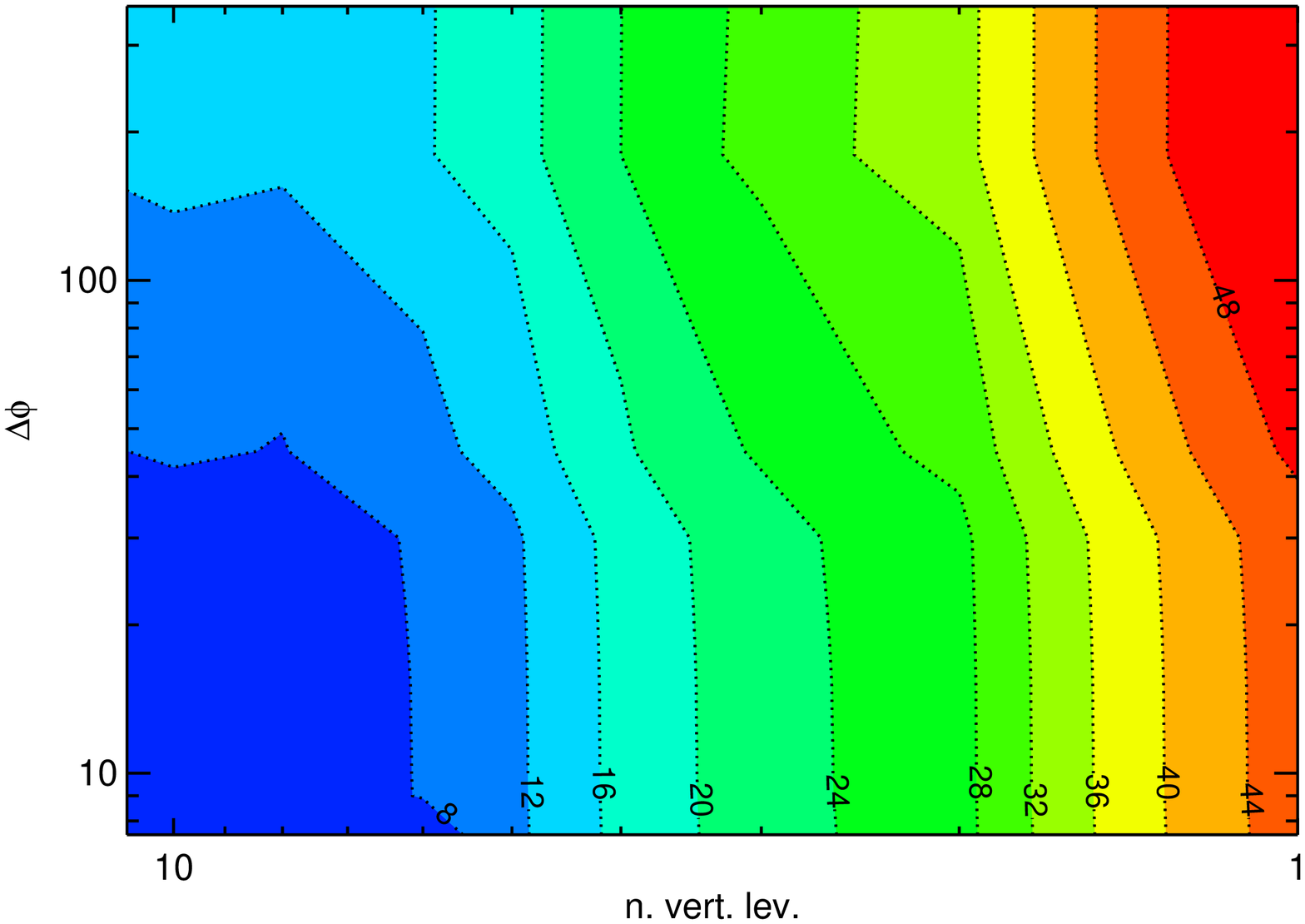}
     \label{vhind3b}
      } 
\caption{Estimates of $\overline{\langle\dot{S}^{ind}_{mat}\rangle_v^\tau}$ (a), and $\Delta\left[\overline{\dot{S}^{ind}_{mat}} \right]_v^\tau$ (b) for $\tau=1$ day, and of  $\overline{\langle\dot{S}^{ind}_{mat}\rangle_v^\tau}$ (c), and $\Delta\left[\overline{\dot{S}^{ind}_{mat}} \right]_v^\tau$ (d) for $\tau=1$ year. The spatial averaging considered here is given by areal averages along horizontal surfaces and vertical averages along columns. The $x$-axis reports the numbers of vertical levels involved in the averaging, the $y$-axis describes the longitudinal extent $\Delta \phi$ of the coarse grained grid boxes. Details are given in the text.    \label{vhind} }
\end{figure}

Finally, we perform spatial averaging along the horizontal and vertical direction and for different values of $\tau$ for the direct and indirect formula of entropy production. In this way, we are able to ascertain the relevance of the processes involving irreversible fluxes across temperature gradients along the vertical direction.  Results are reported in Fig. \ref{vhdir} and Fig. \ref{vhind}, respectively, , and some specific results are given in Table \ref{tabellina3}.   Since we are now dealing with three variables describing the coarse graining - the amplitude in latitude $\Delta \phi$, the number of levels $n$, and $\tau$, we present two cross sections obtained for $\tau=1$ day (virtually indistinguishable from $\tau=1$ hour) and $\tau=1$ year, reported as panels (a) and (c), respectively, in both Figs. \ref{vhdir} and \ref{vhind}. In panels (b) and (d) of these two figures, we report, instead, $\Delta\left[\overline{\dot{S}^{dir}_{mat}} \right]_{v}^\tau$ and  $\Delta\left[\overline{\dot{S}^{ind}_{mat}} \right]_{v}^\tau$, respectively. The inequality $\Delta\left[\overline{\dot{S}^{dir}_{mat}} \right]_{v}^\tau<\Delta\left[\overline{\dot{S}^{ind}_{mat}} \right]_{v}^\tau$ is clearly obeyed.

As we know from the previous discussions, the function $\overline{\langle\dot{S}^{dir}_{mat}\rangle_v^\tau}$ is relatively weakly affected by coarse graining along the horizontal directions and along the time axis. The modest importance of time averaging is confirmed in this more complete analysis, as Figs.  \ref{vhdir2}, and  \ref{vhdir3} are hard to distinguish. Instead, we find that averaging the thermodynamic fields along the vertical reduces  very severely the correlation between the temperature and heating fields, so that the estimate of the entropy production obtained from the coarse grained fields is much smaller than its true value $\overline{\dot{S}_{mat}}$.  The results emphasize that vertical mixing is the dominant effect contributing to the material entropy production.

When considering the indirect formula, we can draw roughly the same conclusions as above, with the difference that $\overline{\langle\dot{S}^{ind}_{mat}\rangle_v^\tau}$ is  more strongly affected by averaging along the horizontal surface and along the time axis. Therefore, Figs.  \ref{vhind2} and  \ref{vhind3} feature clear differences in terms of mean values, and, in each of them, the impact of performing very coarse graining along the horizontal direction is more pronounced with respect to what reported in Fig. \ref{vhdir}. The effect of horizontal coarse graining become noticeable already when grid boxes with side of $(\Delta \lambda,\Delta \phi)\sim (20^\circ,30^\circ)$ are considered. This implies that the spatial-temporal correlation between radiative heating and temperature fields is relevant for larger range of scales than in the case of described above.   These results put in firmer ground the results given in \cite{Luc10}.

Comparing Figs. \ref{vhdir} and \ref{vhind}, one can verify that in all cases $\Delta\left[\overline{\dot{S}^{ind}_{mat}} \right]_{v}^\tau>\Delta\left[\overline{\dot{S}^{dir}_{mat}} \right]_{v}^\tau$. Moreover, one discovers that when considering the coarse possible graining, one obtains that $\overline{\langle\dot{S}^{dir}_{mat}\rangle_{v_{M,v,h}}^{\tau_M}}\sim 16 mWm^{-2}K^{-1}>\overline{\langle\dot{S}^{ind}_{mat}\rangle_{v_{M,v,h}}^{\tau_M}}\sim0$. In order to interpret some these results, we shall consider some limiting cases for Eqs. (\ref{c1})-(\ref{c2}). We first take as averaging volume at each point at surface $v=v_{M,v}$  the vertical column ranging from the bottom of the fluid component of the climate system to the top of the atmosphere, and we consider a long averaging time $\tau=\tau_M \gg 1$ $y$, so that all temporal dependencies are removed. In other terms, we look at the thermodynamic properties of the \textit{ climatological fields}.

We start with the expression relevant for the indirect formula for estimating the material entropy production: 
\begin{equation}
\overline{\langle \dot{S}^{ind}_{mat} \rangle^{\tau_{M}}_{v_{M,v}}}= -\int_V \mathrm{d}^3\mathbf{x}\overline{\left(\frac{\langle \dot{q}_{rad}\rangle_v^\tau}{\langle T\rangle_v^\tau}\right)}= -\int_\Sigma \mathrm{d} {x_1}\mathrm{d}{x_2}\frac{{F_{TOA}({x_1,x_2})}}{{T_{cli}({x_1,x_2})}}.\label{eptransp}
\end{equation}
where given the choice of $\tau$, the time averaging operation given by the overbar in Eq. \ref{eptransp} is immaterial. In Eq. \ref{eptransp}  ${F_{TOA}({x_1,x_2})}=F^{SW}_{TOA}({x_1,x_2})-F^{LW}_{TOA}({x_1,x_2})$ is the climatological average of the net radiative wave flux at the top of the atmosphere (positive when there is net incoming radiation towards the planet), while the lower indices $LW$ and $SW$ indicate the long wave and shortwave components, respectively. ${T_{cli}}({x_1,x_2})$ is the long term mean of the vertical average of the fluid temperature. Such quantity can be closely approximated by the emission temperature ${T_E({x_1,x_2})}=({F^{LW}_{TOA}({x_1,x_2})}/\sigma)^{1/4}$, where $\sigma$ is the Boltzmann's constant. Note that since $F_{TOA}$ and $T_E$ are positively correlated (regions having a net positive incoming radiation are warmer), we have that $F_{TOA}$ and $1/T_E$ are negatively correlated. Since $\int_\Sigma \mathrm{d} {x_1}\mathrm{d}{x_2}F_{TOA}=0$, we derive that $\overline{\langle \dot{S}^{ind}_{mat} \rangle_v^\tau}$ in Eq. (\ref{eptransp}) is positive, as expected.

Equation (\ref{eptransp}) can also be given a different interpretation. We have that ${F_{TOA}({x_1,x_2})}$ is equal to the divergence of enthalpy transport due to the large scale climatological atmospheric and oceanic flow, so that 
\begin{align}
{F_{TOA}({x_1,x_2})}&=\nabla_2\cdot\int_{z_{surf}}^{TOA} \mathrm{d}x_3 [\mathbf{J}_{lat}(\mathbf{x})+\mathbf{J}_{dry}(\mathbf{x})]\nonumber\\&
= \nabla_2\cdot[\mathbf{\tilde{J}}_{lat}(x_1,x_2)+\mathbf{\tilde{J}}_{dry}(x_1,x_2)].
\label{trasportotamarro}
\end{align}
In the previous equation, we have indicated with $\mathbf{{J}}_{lat}(\mathbf{x})=L_w q(\mathbf{x}) \left\{v_1(\mathbf{x}),v_2(\mathbf{x})\right\}$ and $\mathbf{J}_{dry}(\mathbf{x})=[C_p T(\mathbf{x}) +gx_3]\left\{v_1(\mathbf{x}),v_2(\mathbf{x})\right\}$ the large scale, advective horizontal fluxes of latent heat and of dry static energy, respectively, where $L_w$  is the latent heat of evaporation of water (taken as a constant for simplicity), $q$  is the specific humidity, and $C_p$ is the heat capacity of air at constant pressure, and $v_1$ and $v_2$ indicate the two components of the horizontal velocity field. %Instead $\mathbf{{J}}_{pot}(\mathbf{x})=g x_3 \left\{v_1(\mathbf{x}),v_2(\mathbf{x})\right\}$ represents the large scale horizontal advection of potential energy \citep{Peix2,LucariniRagone},  is much smaller than the other terms and can be neglected \cite{Peix2}. 
Finally, the $\tilde{}$  sign refers to the vertically integrated fluxes. Inserting the right hand side of Eq. (\ref{trasportotamarro})  in Eq. (\ref{eptransp}), we conclude that Eq. (\ref{eptransp}) gives the entropy produced by the large scale horizontal transport of the geophysical flows and approximates the climate system as a purely 2D system featuring irreversible heat transport from warm to cold regions.  Looking in the bottom right corner of Fig. \ref{vhind3} (which corresponds to the last entry in Table \ref{tabellina3}), we obtain a value of $\sim6.5$ $mWm^{-2}K^{-1}$ (note that the result is virtually unaltered when averaging over any $\tau\geq 1y$).  

We can bring the previous example to a more extreme case. If the spatial stencil $v$ is the whole climate domain $v_{M,h,v}$, it is easy to derive that:
\begin{equation}
\overline{\langle \dot{S}^{ind}_{mat} \rangle_{v=V}^{\tau_M}} = \mu(v) \frac{\overline{\langle \dot{q}_{rad}\rangle_v^\tau}}{\overline{\langle T\rangle_v^\tau}}=-\mu(v) \frac{\int_\Sigma \mathrm{d} {x_1}\mathrm{d}{x_2}F_{TOA}({x_1,x_2})}{\overline{\langle T\rangle_v^\tau}}=0,\label{c1b}
\end{equation}
because we have reduced the climate to a zero-dimensional system with a unique temperature where absorbed and emitted radiation are equal. Such a system is at equilibrium, cannot do any work, and cannot sustain any irreversible process. See Fig. \ref{vhind3} and penultimate entry in Table \ref{tabellina3}. 

Let's now repeat the same coarse graining operations for the direct formula. We first consider $v=v_{M,v} $. In each location, when integrating vertically, the surface sensible heat fluxes cancel out with the heating rates associated to sensible heat fluxes in the atmosphere, so that for this choice of coarse graining their contribution to the entropy production vanishes. We obtain:
\begin{align}
\overline{\langle \dot{S}^{dir}_{mat} \rangle_{v_{M,v}}^{\tau_M}}&= \int_V \mathrm{d}^3\mathbf{x}\overline{\left(\frac{\langle \dot{q}_{mat}\rangle_v^\tau}{\langle T\rangle_v^\tau}\right)}= \int_\Sigma \mathrm{d} {x_1}\mathrm{d}{x_2}\left[\frac{L_w [P({x_1,x_2})-E({x_1,x_2})]}{{T_{cli}({x_1,x_2})}}+ \frac{{\tilde{\epsilon}^2({x_1,x_2})}}{{T_{cli}({x_1,x_2})}}\right] \nonumber\\&= \int_\Sigma \mathrm{d} {x_1}\mathrm{d}{x_2}\left[\frac{-\nabla_2\cdot \tilde{J}_{lat}({x_1,x_2})}{{T_{cli}({x_1,x_2})}}+ \frac{{\tilde{\epsilon}^2({x_1,x_2})}}{{T_{cli}({x_1,x_2})}}\right].\label{epepsilon}
\end{align}
where $L_w [P({x_1,x_2})-E({x_1,x_2})]=-\nabla_2\cdot \tilde{J}_{lat}({x_1,x_2})$ as imposed by conservation of water  mass, with $P({x_1,x_2})$ and $E({x_1,x_2})$ time-averaged values of precipitation and evaporation, respectively \cite{Peix2}. Furthermore, we indicate with $\tilde{\epsilon}^2({x_1,x_2})$ the vertically  integrated kinetic energy dissipation rate, and we choose, as a first approximation $T_{cli}({x_1,x_2})$ as characteristic temperature defined as before. Therefore, Eq. (\ref{epepsilon}) suggests that the bottom right corner of Fig. \ref{vhdir3} corresponds to the sum of entropy produced by large scale transport of latent heat  plus the entropy produced by the dissipation of kinetic energy.  We find a value of $\sim18$ $mWm^{-2}K^{-1}$. Considering that the entropy production due to large scale transport of sensible heat is much smaller than the corresponding contribution due to latent heat transport (from the precise calculation we get a factor of about 5 as ratio between the two terms) we can derive that the dissipation of kinetic energy contributes for about $\sim13$ $mWm^{-2}K^{-1}$ to the total material entropy.  

%Taking the difference between Eq. (\ref{epepsilon})  and Eq. (\ref{eptransp}), we obtain:
%\begin{equation}
%\overline{\langle S^{dir}_{mat} \rangle_v^\tau}-\overline{\langle S^{ind}_{mat} \rangle_v^\tau}\sim \int_\Sigma \mathrm{d} {x_1}\mathrm{d}x_2\left[\frac{{\tilde{\epsilon}^2({x_1,x_2})}}{{T_{cli}({x_1,x_2})}}+\frac{\nabla_2\cdot \tilde{J}_{dry}({x_1,x_2})}{{T_{cli}({x_1,x_2})}}\right] \label{bias}.
%\end{equation}
%The conjecture presented before suggests that the expression at the r.h.s. of Eq. \ref{bias} is positive. %, which, in fact, is in agreement with the calculations presented in \cite{Luc10}. %Our previous conjecture suggests that this quantity is positive. This is actually the case, as Moreover,  in \cite{Luc10} it was derived that for this choice of coarse graining  $ \overline{\langle S^{dir}_{mat} \rangle_v^\tau} \geq \overline{\langle S^{ind}_{mat} \rangle_v^\tau}$ for any climate, so that the coarse grained direct and indirect formulas are not equivalent. 
%Specifically, using parameters suitable for the present terrestrial climate, we get that Eq. (\ref{eptransp}) accounts for about $15\%$ of the total material entropy production on our planet, while Eq. (\ref{epepsilon}) accounts for about twice than that \citep{Luc10}.; see also the next Section.

Interestingly,  if we compute $\overline{\langle \dot{S}^{dir}_{mat} \rangle_{v_{M,h,v}}^{\tau_M}}$, we do not obtain a vanishing result. While the contribution to entropy production due to heat fluxes is eliminated, the contribution coming from the  dissipation of kinetic energy is not removed by the operation of coarse graining:
\begin{equation}
\overline{\langle \dot{S}^{dir}_{mat} \rangle_{v=V}^\tau} = \mu(v) \frac{\overline{\langle \dot{q}_{mat}\rangle_v^\tau}}{\overline{\langle T\rangle_v^\tau}}=\mu(v) \frac{\int_\Sigma \mathrm{d} {x_1}\mathrm{d}{x_2}\tilde{\epsilon}^{2}({x_1,x_2})}{\overline{\langle T\rangle_v^\tau}}>0.\label{c1c}
\end{equation}
Equation (\ref{c1c}) gives, to a good degree of approximation, the \textit{minimum} value of the entropy production compatible with the presence of a total dissipation $\int_V d^3\mathbf{x}\overline{\epsilon^2(\mathbf{x},t)} $ \citep{Lucarini,Luc10}. The second entry of Table \ref{tabellina3} reports for the contribution given in Eq. \ref{c1c}
 a value of about $16$ $mWm^{-2}K^{-1}$. This value agrees  well with what derived using Eqs. \ref{c1b}-\ref{epepsilon} and with what obtained by direct estimate of $\overline{\dot{S}_{KE}}\sim 13.5$ $mWm^{-2}K^{-1}$.

These results imply that we can obtain an extremely good estimate of the true value of the material entropy production even using climatological, horizontally global  averages of the thermodynamics fields: in such a worst case scenario, we get a bias of about 10\%. Using few selected coarse grained estimates for the entropy production, one can derive that it is possible to split the total value of $52.5$ $mWm^{-2}K^{-1}$ as follows: $\sim6$ $mWm^{-2}K^{-1}$ can be attributed to large scale transports of latent and sensible heat; $\sim13$ $mWm^{-2}K^{-1}$ can be attributed to the entropy produced by dissipation of kinetic energy; the remaining $\sim33.5$ $mWm^{-2}K^{-1}$ can be attributed to vertical transports of sensible and latent heat (basically, convection). These estimates agree quite accurately with what obtained in \citep{Luc10} using scaling analysis. Also, one obtains, in agreement with the inequality proposed in \citep{Luc10}, that the entropy produced by dissipation of kinetic energy is larger than that due to large scale energy transport.

%  Therefore, the physical interpretation of Eq. (\ref{c1c})is rather different from what given in Eq. (\ref{c1b}), even if the coarse graining is the same. 

\section{Conclusions}\label{conclu}
The investigation of the climate system using tools borrowed from non-equilibrium thermodynamics \citep{Pri,Maz} is a very active interdisciplinary research, which allows for connecting concepts of great relevance for climate dynamics, such as large scale heat transports and the Lorenz energy cycle \citep{Peix}, to basic thermodynamical concepts used in the investigation of general non-equilibrium systems \citep{Lucarini,Klei09}. Such a theoretical framework seems relevant especially in the context of the growing field focusing on the study of the atmospheres of exoplanets \cite{Boschi2013,LucAstr}, for which detailed measurements are hardly available. In fact, thermodynamical methods allow for defining inequalities and deducing apparently unexpected relations between different physical quantities \citep{Luc10}. 

%In previous studies, we have emphasized how the material entropy production is an extremely useful thermodynamical indicator of the state of a planet system, and provides, \textit{e.g.}, the clearest indication of whether a planet is a snowball state \citep{LucFr,Boschi2013,LucAstr}. We have also suggested , performing scale analysis over a formula provided by Goody \citep{Goody00}, that it is possible to estimate the total material entropy production and its portion due to large scale horizontal heat transports using low resolution data, and proved the effectiveness of this approach on actual climate models \citep{Luc10}.

In this paper we have focused on understanding where, in the Fourier space, dissipative and irreversible processes are dominant. This has been accomplished by computing how different spatial and temporal scales contribute to the material entropy production in the climate system. We have considered the output coming from a 50 $y$ run under steady state conditions performed with the FAMOUS climate model \citep{FAM}, and have used the entropy diagnostics developed and tested in \citep{Pascale}. We have considered both the direct and the indirect formulas for material entropy production \citep{Goody00}: the former estimates the material entropy production using the heating rates associated to the dissipation of kinetic energy and the convergence of material heat fluxes, the latter uses, instead, the heating rates associated to radiative fluxes.

Our strategy has been the following: we have considered the estimates of the entropy production coming from the coarse grained outputs of the heating and temperature fields. The coarse-graining has been performed both in time and in space. The temporal coarse graining ranges from hourly (timestep of the model) to yearly time scale,  while the spatial coarse grained has been performed in three different modalities: 1) performing longitudinal averages; 2) performing averages along horizontal surfaces; 3) performing mass-weighted averages along the horizontal and vertical directions. We have conjectured and then verified numerically that the coarser the graining of the data, the lower is the resulting estimate of the material entropy production, both in the case of the direct and of the indirect formula for estimating the material entropy production. This implies that at all scales there is a negative correlation between heating rates related to flow (kinetic energy dissipation, sensible and latent heat fluxes) and the temperature field, and a positive correlation between the heating rate due to radiation and the temperature field. In other terms, at all scales, the climate systems results to be forced by radiation, while the resulting forced fluctuations are dissipated by the material fluxes. In agreement with this interpretation, we have conjectured that at all scales the correlation between the radiative heating and the temperature field is stronger than the correlation between the temperature field and the material heating rates. If the two correlation were equal, the climate system would be able to adjust, instantly and locally, to (spatial and temporal) variations in the radiative heating. The numerical results have provided support  for this conjecture.

Considering various special cases of coarse graining, and using the basic thermodynamic equations, we have been able to estimate in a consistent way the contributions to material entropy productions coming from large scale horizontal heat transport ($\sim6$ $mWm^{2-}K^{-1}$), dissipation of kinetic energy ($\sim13$ $mWm^{2-}K^{-1}$), and vertical processes of sensible and latent heat exchanges  (i.e. convection, $\sim33.5$ $mWm^{2-}K^{-1}$). This suggest that, as first approximation, the climate system can be seen in terms of dissipative processes as a collection of weakly coupled vertical columns featuring turbulent exchanges and dissipation. This confirms the ideas presented in \cite{Luc10}.

Note that one could use the quantitative information on the various contributions to the material entropy production to derive some basic properties of the climate system without resorting to the full three dimensional, time dependent fields, In particular, one can derive a good estimate of the intensity of the Lorenz energy cycle by multiplying the estimated value of the contribution of the dissipation to the kinetic energy to the material entropy production times a characteristic temperature of the system, obtaining an estimate of $\sim 3$ $Wm^{-2}$, which is in good agreement with what obtained after processing of the high-resolution data \citep{Pascale}.

The fact that the estimate of the material entropy production of the climate system decreases when a coarser gaining is considered is in qualitative agreement with what derived in the Appendix \ref{app} for the simple case of continuum systems featuring generalized flux-gradient relations. This obviously does not imply that the climate system behaves as a diffusive system, yet share with a diffusive system this interesting property. The fact that at all spatial and temporal scales the system has a positive definite value of material entropy production, when global averages are considered. This is not in contradiction with the well-known phenomena of so-called \emph{negative diffusion}, first noted by Starr \citep{Starr68}, who observed that in certain portions of the atmosphere - namely, near the storm track - the fluxes of momentum transport momentum from low to high momentum regions. While this process - a crucial element of the general circulation of the atmosphere, observed in our model runs as well -  seems to oppose the second law of thermodynamics, it is instead a local but macroscopic phenomenon, where the creation of organized structures (thanks to long-range correlations due to wave propagation), is, as we understand in this paper, over-compensated by large entropy production at the same scales elsewhere in the atmosphere. 

% It is not clear, given a specific model's settings, how relevant these could be, and, indeed, the only way to find this out is to alter the model's resolution. This procedure may have relevance in terms of model tuning, as one could decide to change a model's parameter when altering its resolution in such a way to keep the entropy production constant. 

Apart from providing insights on the properties of forced fluctuations and irreversible dissipative processes in the climate system at various spatial and temporal scales, this paper deals with the relationship between a model, its output, and the chosen observables, by providing information on what is the impact of being able to access data at lower resolution with respect to the model which has generated them. We have learnt that this lack of information always biases negatively our estimate of the entropy production, and that the bias is serious only if we miss information describing the vertical structure of thermodynamic fields. 

Since performing time averages up to the yearly time scale does not bias substantially the estimates of the material entropy production, we have that it is possible to intercompare robustly the state-of-the-art climate models and assess on each of them the impact of climate change on the entropy production by resorting to the output data provided in the freely  accessible PCMDI/CMIP3 (\texttt{http://www-pcmdi.llnl.gov/ipcc/about$\_$ipcc.php}) and PCMDI/CMIP5 (\texttt{http://cmip-pcmdi.llnl.gov/cmip5/}) repositories, where long climate runs outputs are typically stored in the form of monthly averaged data. 

On a different note, the approach presented here seems promising specifically for the investigation of the atmosphere of exoplanets, because it allows for evaluating the error in the estimate of their thermodynamical properties  due to the lack of high-resolution data.

In this paper, we have worked on the post-processing of data. The analysis presented here should be complemented with an additional investigation of how changing the resolution of a model impacts the estimate of its material entropy production, in total, and process by process. Using the material entropy production as cost function for addressing the interplay between the respective role of changes in the resolution of a model and of changes in the coarse graining of the post-processed data seems promising in the tantalizing quest for understanding what is a good model of a geophysical fluid and what is a robust parametrization. The results obtained here seem to have a much more general validity than for the specific case of the present Earth's climate. Therefore, we plan to extend the present analysis, by studying the combined effect of changes in the resolution of the model and in the effective resolution of the post-processed in a simpler geophysical fluid dynamical system like an Aquaplanet. We believe that the conjectures presented on the effect of coarse graining thermodynamic fields on the estimate of the material entropy production is of general validity for a vast range of systems that can be described by continuum mechanics. 

\acknowledgements
The research leading to these results has received funding from the European Research Council under the European Community$'$s Seventh Framework Programme (FP7/2007-2013)/ERC Grant agreement No. 257106. The authors acknowledge the support of the Cluster of Excellence for Integrated Climate Science \textit{CLISAP}.  VL acknowledges the hospitality of the Isaac Newton Institute for Mathematical Sciences (Cambridge, UK) during the 2013 programme \textit{Mathematics for the Fluid Earth}. The authors acknowledge  fruitful scientific exchanges with J. Gregory and the insightful comments of two anonymous reviewers.

\appendix

\section{Spectral Analysis of the Impact of Coarse Graining on the Entropy Production for Diffusive Systems}\label{app}
We wish to provide a simple outlook on how to interpret the results shown in this paper taking a spectral point of view. This is relevant in practical terms because many numerical models of geophysical fluids are actually implemented in spectral coordinates.  We restrict ourselves to the contributions to the material entropy production coming from the presence of material fluxes transporting heat across temperature gradients. 
So, we do not consider here the term responsible for the dissipation of kinetic energy (which is weakly affected by coarse graining) nor the radiative terms contributing to the indirect formula. We consider the case of a much simpler  simpler 3D continuum physical system, where heat transport obeys a generalized diffusive behavior. We make this choice not because we believe that the climate system is, in any real sense, diffusive, but because we wish to show that the observed dependence of the material entropy production on the coarse graining is in qualitative agreement with what would be obtained for a diffusive system.
 
Let's assume that the contribution to the average rate of entropy production coming from the transport of heat due to the flow $J$ across the temperature field $T$ can be computed as:
\begin{equation}
\overline{\dot{S}}^{J}_{mat}=\frac{1}{V}\frac{1}{T}\int_V \mathrm{d}^3\mathbf{x} \int_0^T \mathrm{d}t\vec{J}\cdot\vec{\nabla}\frac{1}{T}=-\frac{1}{V}\frac{1}{T}\int_V \mathrm{d}^3\mathbf{x} \int_0^T \mathrm{d}t \frac{\vec{J}\cdot\vec{\nabla} T}{T^2}=-\frac{1}{V}\frac{1}{T}\int_V \mathrm{d}^3\mathbf{x} \int_0^T \mathrm{d}t \frac{\vec{\nabla}\cdot \vec{J}}{T}
\end{equation}
Let's now make the simplifying diffusive-like assumption that $\vec{J}=-\vec{\nabla}G(T)$, where $dG/dT>0$, so that the flux is always opposed in verse to the temperature gradient; in the usual linear flux-gradient approximation we have $G(T)=\kappa T$, $\kappa>0$. We derive: 
\begin{equation}
\overline{\dot{S}}^{J}_{mat}=\frac{1}{V}\frac{1}{T}\int_V \mathrm{d}^3\mathbf{x} \int_0^T \mathrm{d}t\ \frac{G'(T)}{T^2}|\vec{\nabla} T|^2=\frac{1}{V}\frac{1}{T}\int_V \mathrm{d}^3\mathbf{x} \int_0^T \mathrm{d}t\ |\vec{\nabla} \Psi(T)|^2>0
\end{equation}
where $\Psi(T)=\int dT \sqrt{G'(T)/T^2}$. For sake of simplicity - but without loss of generality - we assume that our domain $\Sigma$ is a parallelepiped of sides $L_x$, $L_y$, and $L_z$. Using Parseval's theorem, we derive that the rate of entropy production can be written as:       
\begin{equation}
\overline{\dot{S}}^{J}_{mat}=\sum_{p,q,r,s}(k_p^2+k_q^2+k_r^2) |\Psi_{p,q,r,s}|^2= \sum_{p,q,r,s}S^{J,mat}_{p,q,r,s}
\label{fourier}
\end{equation}
where
\begin{equation}
\Psi_{p,q,r,s}=\frac{1}{V}\frac{1}{T}\int_V \mathrm{d}^3\mathbf{x} \int_0^T \mathrm{d}t\ \Psi\exp[2\pi i(p/L_x x+q/L_y y+r/L_z z-s/T t)].
\end{equation}
 Performing a spatio-temporal coarse graining to the field $\Psi\rightarrow\tilde{\Psi}$ can be reframed as applying a linear filter in Fourier space, as $\Psi_{p,q,r,s}\rightarrow\tilde{\Psi}_{p,q,r,s}=\Phi_{p,q,r,s}\Psi_{p,q,r,s}$, where $|\Phi_{p,q,r,s}|\leq 1$ $\forall p,q,r,s$ plus the usual complex conjugacy properties. Note that in our case we would like to able to apply the filtering to the $T$ field and to the heating field $-\vec{\nabla}\cdot\vec{J}$ and not to the function $\Psi$ constructed here. Nonetheless, assuming that the relative change of $T$ across the domain (see also Eq. (\ref{d1})) is small, the conclusions are virtually unaltered.  
 
Equation \ref{fourier} has the remarkable property that all of terms of the summation $S^{J,mat}_{p,q,r,s}$ are positive. We can interpret $S^{J,mat}_{p,q,r,s}$ as the entropy produced by processes occurring at the scales described by the indices, in this case $\lambda_x=L_x/p$, $\lambda_y=L_y/q$, $\lambda_z=L_z/r$, and $\tau=T/s$.
 Therefore, indicating as $\tilde{\overline{\dot{S}}}^{J}_{mat}$ the value of the entropy production for the coarse grained fields, we obtain:
\begin{equation}
\tilde{\overline{\dot{S}}}^{J}_{mat}=\sum_{p,q,r,s}(k_p^2+k_q^2+k_r^2) |\Phi_{p,q,r,s}|^2|\Psi_{p,q,r,s}|^2=\sum_{p,q,r,s}|\Phi_{p,q,r,s}|^2 S^{J,mat}_{p,q,r,s}\leq \overline{\dot{S}}^{J}_{mat}.
\label{fourier2}
\end{equation}
In particular, we can associate a coarse graining on the scales $\Lambda_x$, $\Lambda_y$, $\Lambda_z$, and $\tau$ (referred to the $x$-, $y$-, $z$-directions and time, respectively) to a filter of the form 
$\Phi_{p,q,r,s}=0$ if $p>L_x/\Lambda_x$, or $q>L_y/\Lambda_y$, or $r>L_z/\Lambda_z$, or $s>T/\tau$ and $\Phi_{p,q,r,s}=1$ otherwise. Slightly different ways of doing the coarse graining will result into different filters, which will be, nonetheless, asymptotically equivalent if the involved scales are the same. 

The main conceptual point behind this result is independent of the shape of the of the domain of integration: the natural orthogonal expansion for atmospheric fields defined in an (approximately) spherical thin shell is given by spherical harmonics in for the latitudinal and longitudinal dependence and the usual Fourier expansion for the vertical direction. In the case of a thin spherical shell of thickness $L_z$ situated at distance $R$ from the center of the sphere, Eq. \ref{fourier} can be rewritten as:  
\begin{equation}
\overline{\dot{S}}^{J}_{mat}=\sum_{n}\sum_{l\geq 0}\sum_{m=-l}^l\sum_{s}(k_n^2+l(l+1)/R^2) |\Psi_{n,l,m,s}|^2
\label{spec}
\end{equation}
with
\begin{equation}
\Psi_{n,l,m,s}=\frac{1}{L_z}\frac{1}{4\pi}\frac{1}{T}\int_\Omega d\Omega \int_0^T  \Psi(z,\theta,\phi,t)\exp[2\pi i(n/L_z z-s/T t)]*Y(\theta,\phi)_n^{m*}.
\label{sph}
\end{equation}
where $Y(\theta,\phi)_n^{m}$ are the usual spherical harmonics and $\Omega$ refers to the solid angle. In this case, performing a spatio-temporal coarse graining to the field $\Psi\rightarrow\tilde{\Psi}$ results in reduced value of the estimate of the entropy production:
\begin{equation}
\tilde{\overline{\dot{S}}}^{J}_{mat}=\sum_{n}\sum_{l\geq 0}\sum_{m=-l}^l\sum_{s}(k_n^2+l(l+1)/R^2) |\Phi_{n,l,m,s}|^2|\Psi_{n,l,m,s}|^2\leq \overline{\dot{S}}^{J}_{mat}.
\label{spec2}
\end{equation}
It is now easy to relate the expression of $|\Phi_{n,l,m,s}|^2$  to common averaging operation performed on climate data. A coarse graining on the vertical scale $\Lambda_z$, on a temporal scale $\tau$ and on a horizontal surface area $\sigma$ (or angular resolution $\sigma/R^2$) amounts to setting $|\Phi_{n,l,m,s}|^2=0$ if $l\geq\sqrt{8\pi R^2/\sigma}$ (corresponding approximately to a triangular truncation $T(K)$, where $K$ is the integer closest to $\sqrt{8\pi R^2/\sigma}$), or $n>L_z/\Lambda_z$, or $s>T/\tau$, and $|\Phi_{n,l,m,s}|^2=1$ otherwise. Instead, performing zonal averages corresponds to setting, $|\Phi_{n,l,m,s}|^2=0$ if $m\neq 0$. 

The bottom line of the previous considerations is that  adopting a coarser graining corresponds to increasing the involved scales determining the spectral cutoff. if we assume a flux-gradient relationship which is consistent with the second law of thermodynamics (even if it is not the usual Fickian, linear relation), Eqs. \ref{fourier2} and \ref{spec2} imply that as the graining becomes coarser, the estimate of the entropy production becomes smaller, because we the summation is performed over fewer terms, all of them positive. This behavior is independent of the physical domain under consideration. Moreover, the previous considerations qualitatively apply - even if results are somewhat more cumbersome - if the relationship between flux and gradient is more general than what previously assumed, e.g. if  $J_i=-\partial_iG_i(T)$ (where the Einstein summation convention is not taken), under the condition that $dG_i/dT<0$ $\forall i$.

\providecommand{\bysame}{\leavevmode\hbox to3em{\hrulefill}\thinspace}
\providecommand{\MR}{\relax\ifhmode\unskip\space\fi MR }
\providecommand{\MRhref}[2]{%
  \href{http://www.ams.org/mathscinet-getitem?mr=#1}{#2}
}
\providecommand{\href}[2]{#2}

\newpage

\begin{table}[htdp]
\caption{List of the main FAMOUS' parametrization routines for unresolved processes and their impact in term of heating rates $\dot{q}_k^c$ on the various terms contributing to $\dot{s}_{mat}$. Codes: BL=Boundary Layer; AC=Atmospheric Convection: HD=Hyperdiffusion; OC=Oceanic Convection; D=Diffusion; GW=Gravity Waves; C/E=Condensation/Evaporation: ML=Mixed Layer}
\begin{center}
\begin{tabular}{|c|c|c|c|c|}
Contribution to $\dot{s}_{mat}$  & Atmosphere & Ocean & Soil & Cryosphere \\
\hline
$\frac{\epsilon^2}{T}$ & BL, C, GW, HD & & &\\  
\hline
$-\frac{\nabla \cdot \mathbf{F_{LH}}}{T}$ & BL, AC, C/E, & BL & BL & BL\\
\hline
$-\frac{\nabla \cdot  \mathbf{F_{SH}}}{T}$ & BL, AC, HD & BL, ML, OC, D & BL & BL\\
\hline
\end{tabular}
\end{center}
\label{proce}
\end{table}

\begin{table}[ht]
\caption{Values of  $\Delta\left[\overline{\dot{S}^{dir}_{mat}} \right]^\tau_{v}$ and  $\Delta\left[\overline{\dot{S}^{ind}_{mat}} \right]^\tau_{v}$ obtained when considering the coarsest resolution in space (lower index: $v_{M}$), in time (lower index: $\tau_{M}$), or both. Only 2D horizontal averagings are considered here.  Reference values for highest resolution data: $\overline{\dot{S}_{mat}^{dir}}=52.5$ $mW m^{-2} K^{-1}$ and $\overline{\dot{S}_{mat}^{ind}}=52.1$ $mW m^{-2} K^{-1}$. All  values are in units of $mW m^{-2} K^{-1}$. \label{tabellina2}}
\centering
\begin{tabular}{l c c c c c c c c c c c c c c  }
\hline
Averaging                     & $\Delta\left[\overline{\dot{S}^{dir}_{mat}} \right]_{v_{M}}$     &  $\Delta\left[\overline{\dot{S}^{dir}_{mat}} \right]^{\tau_{M}}_{v_{M}}$  &    $\Delta\left[\overline{\dot{S}^{dir}_{mat}} \right]^{\tau_{M}}$ & $\Delta\left[\overline{\dot{S}^{ind}_{mat}} \right]_{v_{M}}$     &  $\Delta\left[\overline{\dot{S}^{ind}_{mat}} \right]^{\tau_{M}}_{v_{M}}$  &    $\Delta\left[\overline{\dot{S}^{ind}_{mat}} \right]^{\tau_{M}}$\\
\hline
                     \hline
Longitudinal       &2.1  &  2.2 & 2.1 & 2.2 & 5.0&5.0 \\
\hline 
Surface &  5.3 & 5.3 & 2.2& 12.3 & 12.8 & 5.0 \\ 
\hline
%Angular  res.          &           \\
%$\dot{S}_{mat}^{dir}$           & 5.3 & 5.3  &  2.2\\
%$\dot{S}_{mat}^{ind}$ & 13 &  13     & 4.8\\
%$\dot{S}_{hyper}$ &  0.7 & 0.7 & 0.9 \\
%\hline
%Equi-area res.        &   \\
%$\dot{S}_{mat}^{dir}$           &5.1  & 5.1  & 1.9 \\
%$\dot{S}_{mat}^{ind}$ &   13&    12.9 &4.9 \\
%$\dot{S}_{hyper}$ & 0.5&  0.5 &0.5 \\
%\hline
\end{tabular}
\end{table}

\begin{table}[ht]
\caption{Values of  $\Delta\left[\overline{\dot{S}^{dir}_{mat}} \right]^\tau_{v}$ and  $\Delta\left[\overline{\dot{S}^{ind}_{mat}} \right]^\tau_{v}$ obtained when considering the coarsest resolution either in horizontal direction (lower index: $v_{M,h}$ or vertical direction (lower index: $v_{M,v}$), or in both  (lower index: $v_{M,h,v}$). $\tau$ is set to $1$ y. Reference values for highest resolution data: $\overline{\dot{S}_{mat}^{dir}}=52.5$ $mW m^{-2} K^{-1}$ and $\overline{\dot{S}_{mat}^{ind}}=52.1$ $mW m^{-2} K^{-1}$. All  values are in units of $mW m^{-2} K^{-1}$.  \label{tabellina3}}
\centering
\begin{tabular}{l c c c c c c c c c c c c c c  }
\hline
Averaging                     & $\Delta\left[\overline{\dot{S}^{dir}_{mat}} \right]^{\tau_M}_{v_{M,h}}$     &  $\Delta\left[\overline{\dot{S}^{dir}_{mat}} \right]^{\tau_{M}}_{v_{M,h,v}}$  &    $\Delta\left[\overline{\dot{S}^{dir}_{mat}} \right]^{\tau_{M}}_{v_{M,v}}$ & $\Delta\left[\overline{\dot{S}^{ind}_{mat}} \right]^{\tau_M}_{v_{M,h}}$     &  $\Delta\left[\overline{\dot{S}^{ind}_{mat}} \right]^{\tau_{M}}_{v_{M,h,v}}$  &    $\Delta\left[\overline{\dot{S}^{ind}_{mat}} \right]^{\tau_{M}}_{v_{M,v}}$\\
\hline
\hline
Mass weighted       &5.3  &  36.5 & 34.5 & 12.8 & 52.1& 45.6 \\
%\hline 
%Areall &  5.3 & 5.3 & 2.2& 12.3 & 12.8 & 5.0 \\ 
\hline
%Angular  res.          &           \\
%$\dot{S}_{mat}^{dir}$           & 5.3 & 5.3  &  2.2\\
%$\dot{S}_{mat}^{ind}$ & 13 &  13     & 4.8\\
%$\dot{S}_{hyper}$ &  0.7 & 0.7 & 0.9 \\
%\hline
%Equi-area res.        &   \\
%$\dot{S}_{mat}^{dir}$           &5.1  & 5.1  & 1.9 \\
%$\dot{S}_{mat}^{ind}$ &   13&    12.9 &4.9 \\
%$\dot{S}_{hyper}$ & 0.5&  0.5 &0.5 \\
%\hline
\end{tabular}
\end{table}

\newpage

% \begin{figure}
% \centering
   
%\caption{ 
%\label{fig2}} 
%\end{figure}

% \begin{figure}
% \centering
% \subfigure[]{
  %  \includegraphics[angle=-180, width=0.8\textwidth]{deriv.pdf}
%    \label{der1}
 %  }
%   \subfigure[]{
 %    \includegraphics[angle=-180, width=0.8\textwidth]{deriv2.pdf}
 %    \label{der2}
 %     } 
%\caption{                                        \label{der}  }
%\end{figure}

 %\begin{figure}
% \centering
% \subfigure[]{
   % \includegraphics[angle=-0, width=0.5\textwidth]{verthor_3h.pdf}
  %  \label{vhind1}
  % }
  % \subfigure[]{
  %   \includegraphics[angle=-0, width=0.5\textwidth]{verthor_day.pdf}
  %   \label{vhind2}
  %    } 
 %  \subfigure[]{
   %  \includegraphics[angle=-0, width=0.5\textwidth]{verthor_year.pdf}
  %   \label{vhind3}
  %    } 
%\caption{  \label{vhind} }
%\end{figure}

% \begin{figure}
 %\centering
 %\subfigure[]{
   % \includegraphics[angle=-0, width=0.46\textwidth]{verthor_zon_3h.pdf}
   % \label{}
  % }
   %\subfigure[]{
   %  \includegraphics[angle=-0, width=0.46\textwidth]{verthor_zon_day.pdf}
    % \label{}
    %  } 
  % \subfigure[]{
   %  \includegraphics[angle=-0, width=0.46\textwidth]{verthor_zon_year.pdf}
    % \label{}
    %  } 
%\caption{  \label{} }
%\end{figure}

% \begin{figure}
% \centering
% \subfigure[]{
%    \includegraphics[angle=-0, width=0.46\textwidth]{verthor_dir_zon_3h.pdf}
   % \label{}
  % }
 %  \subfigure[]{
  %   \includegraphics[angle=-0, width=0.46\textwidth]{verthor_dir_zon_day.pdf}
    % \label{}
     % } 
 %  \subfigure[]{
  %   \includegraphics[angle=-0, width=0.46\textwidth]{verthor_dir_zon_year.pdf}
   %  \label{}
  %    } 
%\caption{  \label{} }
%\end{figure}

\end{document}